\documentclass[10pt]{iopart}

\usepackage{graphicx}
\usepackage[usenames]{color}
 
        \usepackage{amssymb}

        \usepackage[usenames]{color}
        \usepackage[normalem]{ulem}

\begin{document}


        \title{Self-trapped quantum balls  in  binary Bose-Einstein condensates}

        \author{Sandeep Gautam\footnote{sandeep@iitrpr.ac.in}}
        \address{Department of Physics, Indian Institute of Technology Ropar, Rupnagar, Punjab 140001,
                     India
                     }
        \author{S. K. Adhikari\footnote{adhikari44@yahoo.com,
                  http://www.ift.unesp.br/users/adhikari}}
        \address{Instituto de F\'{\i}sica Te\'orica, Universidade Estadual
                     Paulista - UNESP,  01.140-070 S\~ao Paulo, S\~ao Paulo, Brazil}


        \date{\today}
     \begin{abstract}

We study the formation of a  stable self-trapped spherical quantum ball in a binary Bose-Einstein 
condensate (BEC) with two-body inter-species attraction and intra-species repulsion 
employing  the beyond-mean-field Lee-Huang-Yang  and the 
three-body interactions. We find that either of these interactions  or 
a combination of these  can stabilize the binary BEC quantum ball with very similar 
stationary results, and for a complete description of the problem both the
terms should be considered.  These interactions lead to higher-order nonlinearities, e.g. quartic and
quintic, respectively, in a nonlinear dynamical equation compared to the cubic nonlinearity 
of the two-body contact interaction in the mean-field Gross-Pitaevskii equation. The higher-order 
nonlinearity makes the energy infinitely large at the center of the binary ball and thus avoids 
its collapse.  In addition to the formation of stationary
binary  balls, we also study a collision between two such balls.  At large velocities, 
the collision is found to be elastic, which turns out to be inelastic as the velocity is lowered.  
We consider the numerical solution of a  beyond-mean-field model for the binary ball   
as well as a single-mode variational approximation to it in this study. 

\end{abstract}

        \maketitle
 \section{Introduction}
\label{Sec-I}
A one-dimensional (1D) matter-wave bright soliton, bound due to a balance 
between nonlinear attraction and   defocusing forces, can travel at a 
constant velocity \cite{1D}. Solitons have been studied and observed in 
different classical and quantum systems, such as, on water surface, and 
in nonlinear optics \cite{nlo} and Bose-Einstein condensate (BEC) 
\cite{expt}. In many cases, the 1D soliton is found to be 
analytic with momentum and energy conservation which guarantee elastic 
collision between solitons with shape preservation. In a BEC, 
quasi-1D solitons have been realized \cite{expt} in a cigar-shaped configuration 
with strong confining traps in transverse directions following a theoretical 
suggestion \cite{theo}.  However, such a soliton cannot be realized \cite{1D,nlo}
in a three-dimensional (3D) system due to a collapse instability for attractive 
interaction. 

    The theoretical studies on BEC solitons are usually based on the  
mean-field Gross-Pitaevskii (GP) equation \cite{gp} with two-body contact interaction. 
It has been  demonstrated that an inclusion of the beyond-mean-field 
Lee-Huang-Yang (LHY) interaction \cite{Lee, Petrov1} or of a three-body interaction 
\cite{adhikari} in the dynamical model, both leading to a higher-order repulsive 
nonlinear term compared to the cubic nonlinear two-body attraction 
in the  GP equation, can arrest the collapse and lead to a self-bound BEC droplet. 
Petrov \cite{Petrov1} demonstrated the formation of a binary BEC droplet for 
an intra-species repulsion with LHY interaction and an inter-species attraction. 
One of us \cite{adhikari} demonstrated the formation of a BEC quantum ball for 
an attractive two-body and a repulsive three-body interaction.
We prefer the name quantum ball or simply ball  over quantum {\it droplet} for the localized
{\it  nondipolar} BEC state after establishing the 
robustness of such a  state to maintain the spherical ball-like shape after collision 
\cite{adhikari} in contrast to easily deformable liquid droplets.
 A binary 
boson-fermion quantum ball can be formed for an  attractive boson-fermion and 
repulsive boson-boson  interaction together   
with the LHY interaction and/or  with a repulsive three-boson interaction  \cite{arxiv}, while the fermions remain 
quasi-noninteracting.  A {quantum} ball can also be formed in a multi-component spinor 
BEC with spin-orbit or Rabi coupling \cite{malomed}. 
{
The spin-orbit coupling and Raman coupling can also generate an effective 
interatomic repulsion cancelling the mean-field attraction and stabilizing the binary trapped condensates 
against collapse \cite{Zi-Fa}. The exchange induced spin-orbit coupling, when only one
of the component of the binary system is coupled with Raman lasers, has recently been
studied \cite{Chen}.}  {
 There also has been suggestion for dynamically stabilized self-bound 
3D states \cite{AD}.} 
Quantum droplets were  realized and studied in     dipolar $^{164}$Dy 
\cite{Kadau}  and 
$^{166}$Er \cite{Chomaz}
{BECs}.   The formation of dipolar  droplets
was later explained by including {an} LHY  interaction \cite{Wachtler1,Schmitt,Barbut,Bisset,Wachtler2} 
or a three-body interaction \cite{K3} to the  contact interaction.
Following the suggestion of Petrov \cite{Petrov1}, more recently, a  binary BEC ball 
of two hyperfine states of $^{39}$K for attractive inter-species and repulsive 
intra-species interactions has been observed \cite{Cabrera,Cheiney,inguscio}.

In the previous studies of quantum droplets in dipolar  
\cite{Wachtler1,Barbut,Bisset,Wachtler2} and binary  \cite{Cabrera,Cheiney,inguscio} BECs,   
the LHY interaction was considered to be responsible for binding. The LHY interaction leads 
to a higher-order repulsive quartic nonlinearity in the dynamical model compared to the 
cubic nonlinearity of the GP equation arising from the two-body contact interaction in the Hartree approximation. 
In fact, any higher-order repulsive nonlinearity, or their combination, can stabilize 
the droplet.

In this study, we consider a repulsive three-body interaction  together 
with the LHY interaction for the   formation of a stable self-bound spherical binary BEC  
quantum ball with repulsive intra-species and attractive inter-species interaction. 
We find that either the LHY interaction or a repulsive three-body interaction or a combination 
of both can stabilize the binary ball with very similar {results}, and for 
a complete description of the problem both the terms should be considered.  Without these higher-order 
terms, an attractive BEC has an infinite negative  energy at the center leading to 
a collapse of the system to the center. Any higher-order repulsive term, however small it 
may be, leads to an infinite positive energy at the center and stops the collapse.

  We consider a binary $^{87}$Rb system in two different hyperfine states.  We derive  the binary model
equations with the LHY interaction and a repulsive three-body term and solve it 
numerically without approximation. In addition, we consider a variational approximation {\cite{vari}}  to 
our model in a single-mode approximation (SMA) when the binary set of equations is reduced to a single 
equation. The SMA is commonly used \cite{spin} in a description of a spinor BEC. 
Two different variational {\em ansatz} $-$ a Gaussian and a modified Gaussian $-$  were used  for a better 
approximation to the density profile. In addition to the formation of stationary binary  balls, 
we also  considered  moving balls and collisions between two such balls. At high velocities, 
the collision was found to be essentially elastic with practically no deformation. At lower velocities,  
the collision turns inelastic.  

In Sec. \ref{IIA} we derive the  binary model nonlinear Schr\"odinger (NLS) equations 
appropriate for this study. To make these equations, with complicated nonlinear terms, 
analytically tractable, we consider a SMA   in Sec. \ref{IIB}. A Gaussian variational 
approximation is then implemented on the model SMA equation.  Numerical results  of the model using the split time-step 
Crank-Nicolson \cite{Muruganandam} and Fourier pseudo-spectral \cite{Bao}  methods for stationary binary balls
are  presented   
in Sec. \ref{IIIA}  and compared with  variational results. In Sec. \ref{IIIB} the numerical results for collision 
dynamics of two binary balls are considered. 
 A summary and discussion of the study {is} presented in Sec. \ref{IV}.

\section{Analytical Result}

\label{II}

\subsection{Binary model equations}

\label{IIA}
 
The interaction  energy density ${\cal E}$ (energy per unit volume) of a homogeneous dilute weakly repulsive Bose gas with 
LHY interaction \cite{Lee}  as well as  two and  three-body interactions is given by {\cite{Petrov1,adhikari}} 
\begin{eqnarray}
{\cal E} &=& \frac{U n^2}{2}\left( 1+\frac{128\sqrt{na^3}}{15\sqrt{\pi}}\right) + K_3 \frac{n^3}{6},\label{EbyV}\\
            &=& \frac{U n^2}{2} + \frac{8m^4}{15\pi^2\hbar^3}c^{5} +K_3 \frac{n^3}{6} 
\label{E/V},
\end{eqnarray}
where $n$ is the number density, the two-body interaction strength $U = 4\pi\hbar^2a/m$,   $a$ is the s-wave scattering
length, $m$ is the mass of an atom,   $K_3$ is the three-body interaction strength, $c = \sqrt{Un/m}$ is the 
speed of sound in the single component BEC \cite{gp,Pethick}.
The first two terms on the right-hand-side (rhs) of (\ref{E/V}) are the two-body interaction and its LHY correction, respectively, and the last term is the three-body interaction.  
The LHY interaction  is equal to the zero point energy
of the Bogoliubov modes \cite{Petrov1}.   
The bulk chemical potential $\mu \equiv \partial {\cal E}/\partial n$ is the nonlinear interaction term of 
the following time-dependent mean-field NLS equation for a trapped BEC:  
\begin{eqnarray} 
{\mathrm i}\hbar\frac{\partial \psi}{\partial t} & = &\frac{-\hbar^2 \nabla^2}{2m}\psi + V\psi + U|\psi|^2\left(1 + \frac{32 a^{3/2} |\psi|}{3\sqrt{\pi}}\right)\psi 
                                       +  \frac{K_3}{2} |\psi|^4\psi,
\end{eqnarray}
where $V$ is the trapping potential,  $\psi({\bf r},t)$ is the 
condensate wave function and, in terms of it, condensate density $n({\bf r}) = |\psi({\bf r})|^2$ 
with the normalization $\int |\psi|^2 d{\bf r}=N$, where $N$ is the number of trapped atoms in the BEC.

The beyond-mean-field LHY energy contribution to energy, viz.   (\ref{E/V}), has limited validity 
for only small scattering lengths \cite{PLA}.  For larger scattering lengths, specially at unitarity as $a\to \infty$, this term diverges even faster than the GP term proportional to scattering length, while the energy density should be finite. An analytic beyond-mean-field energy density valid for both small and large scattering lengths has been given \cite{AS}. However, for values of scattering lengths considered in this study the LHY expression (\ref{E/V}) gives the actual state of affairs.

In a binary BEC, there are two speeds of sound  $c_i$ \cite{Pethick}, where suffix $i=1,2$ identifies the two species of atoms,  
and the  energy density with LHY and three-body interactions in a homogeneous medium is 
\cite{Eckart}
\begin{eqnarray}
{\cal E} &=& \sum_i \left[\frac{U_in_i^2}{2}  + \frac{8m^4}{15\pi^2\hbar^3} c_i^5 \right]   + U_{12}n_1n_2  
             + \frac{K_3}{6} n^3,
\end{eqnarray}
where $n=n_1+n_2.$
Here intra-species and inter-species three-body interaction strengths and  masses  are
taken to be  equal to $K_3$ and $m$, respectively, the two-body intra- and inter-species interaction strengths 
are  $U_i = 4\pi\hbar^2a_i/m$ with $a_i$  the $s$-wave intra-species scattering length for species $i$, and 
$ U_ {12} = {4\pi \hbar^2 a _{12} /m}$,
where $a_{ 12}$  is  the $s$-wave inter-species scattering length, respectively.
  The two speeds of the sound for the system are \cite{Pethick,Gladush}
\begin{eqnarray}
{c_{\pm} = \sqrt{\frac{\sum_iU_in_i\pm\sqrt{(U_1n_1-U_2n_2)^2+4n_1n_2U_{12}^2}}{2m}}.} 
\end{eqnarray}
If $U_{12} \approx -\sqrt{U_1U_2}$ for repulsive intra-species and attractive inter-species interactions, 
${c_-}\approx 0$ and ${c_+} =\sqrt{(U_1n_1+U_2n_2)/m}$. Then, the mean field energy with LHY and three-body interactions 
can be written as
\begin{eqnarray}
{\cal E} & =& \sum_i \frac{U_in_i^2}{2}  + U_{12}n_1n_2 + \frac{8m^4}{15\pi^2\hbar^3}\left( \frac{{\sum_i} U_in_i}{m}\right)^{5/2} 
             + \frac{K_3}{6} n^3 \label{E/V_bin}.
\end{eqnarray}
In this case the chemical potentials $\mu_i \equiv  \partial{\cal E}/\partial n_i$ are the nonlinear terms of the following  binary time-dependent NLS 
equation for the localized  BEC mixture 
\begin{eqnarray}
{\mathrm i}\hbar\frac{\partial \psi_1}{\partial t}  &= &\biggr[\frac{-\hbar^2 \nabla^2}{2m} +V+ U_1|\psi_1|^2 + U_{12}|\psi_2|^2+\frac{K_3}{2}\Big(\sum_i |\psi_i|^2\Big)^2 \nonumber
\\
&+& 
\frac{32U_1 {\left(\sum_i a_i |\psi_i|^2 \right)}^{3/2}}{3\sqrt{\pi}}\biggr]\psi_1,\label{GPE1}
\\
{\mathrm i}\hbar\frac{\partial \psi_2}{\partial t}  &= &\biggr[\frac{-\hbar^2 \nabla^2}{2m} +V+ U_2|\psi_2|^2 + U_{12}|\psi_1|^2+ \frac{K_3}{2}\Big(\sum_i |\psi_i|^2\Big)^2 \nonumber \\&+&
\frac{32 U_2{\left(\sum_i a_i |\psi_i|^2 \right)}^{3/2}}{3\sqrt{\pi}}\biggr]\psi_2,\label{GPE2}
\end{eqnarray}
with $n_i({\bf r}) =|\psi_i({\bf r})|^2$ and $\int |\psi_i|^2d{\bf r} =N_i$, where $N_i$ is the number of atoms in species $i$.

Let us define $l_0 \equiv 1$ $ \mu$m as a scaling length which we use  to rewrite the  (\ref{GPE1})-(\ref{GPE2})
in  dimensionless form. To this end, we  write length, number density, time, and energy 
in the units of $l_0$, $l_0^{-3}$, $ml_0^{2}/\hbar$, and $\hbar^2/ml_0^{2}$, respectively. The dimensionless
version of  (\ref{GPE1})-(\ref{GPE2}) thus obtained is 
\begin{eqnarray}
{\mathrm i} \frac{\partial \phi_1}{\partial \tilde{t}}  &= &\biggr[\frac{- \tilde{\nabla}^2}{2}+\tilde V + \tilde{U}_1|\phi_1|^2 + \tilde{U}_{12}|\phi_2|^2+\frac{\tilde{K}_3}{2}\Big(\sum_i N_i |\phi_i|^2\Big)^2 \nonumber \\ &+& 
 \frac{16 \tilde{a}_1 {\left(\sum_i U_i |\phi_i|^2 \right)}^{3/2}}{3{\pi}}\biggr]\phi_1,\label{GPEs1}\\
{\mathrm i} \frac{\partial \phi_2}{\partial \tilde{t}}  &= &\biggr[\frac{-\tilde{\nabla}^2}{2} +\tilde V+ \tilde{U}_2|\phi_2|^2 + \tilde{U}_{21}|\phi_1|^2+ \frac{\tilde{K}_3}{2}\Big(\sum_i N_i |\phi_i|^2\Big)^2
\nonumber \\  &+&
 \frac{16 \tilde{a}_2 {\left(\sum_i U_i |\phi_i|^2 \right)}^{3/2}}{3{\pi}}\biggr]\phi_2,\label{GPEs2}
\end{eqnarray}
where the   dimensionless   variables are defined by  $|\phi_i|^2 = N_i^{-1}{|\psi_i|^2} l_0^3, \tilde{U}_i= 4\pi N_i \tilde{a}_i,  \tilde{U}_{12}= 4\pi N_2 \tilde{a}_{12},  \tilde{U}_{21} 
= 4\pi N_1 \tilde{a}_{12}, \tilde{K}_3 =m K_3 /(\hbar l_0^4). $
The normalization conditions satisfied by the dimensionless wave functions now are
$\int |\phi_i|^2 d\tilde{\bf r} \equiv \int \tilde n_i d\tilde{\bf r} = 1$.
To simplify  the notation,  we will denote the dimensionless
variables without tilde except if stated otherwise. All the symbols used in the rest of the paper  are 
dimensionless.  In the following, for the formation of a self-trapped binary ball  we will set the trapping potential $V=0$.

\subsection{Single-Mode Variational Approximation}
\label{IIB}
 
The Lagrangian density  of the spherical  binary ball with LHY   and  three-body interactions   is 
\begin{eqnarray}
{\cal L}& = &   {\sum_i  \frac{N_i}{2} \Big\{  {\mathrm i}  \big(  \phi_i   \phi_i^{*'} - \phi_i^*
 \phi_i ^{'}\big)+  \big| \nabla \phi_i\big|^2   
}
{+U_i|\phi_i|^4\Big\}} +N_1U_{12}|\phi_1|^2 |\phi_2|^2
\nonumber \\
& 
+& \frac{8(\sum_iU_i|\phi_i|^2)^{5/2}}{15\pi^2} + \frac{K_3 \big(\sum_i N_i|\phi_i|^2\big)^3 }{6} ,
 \label{Lagrangian}
\end{eqnarray}
where  prime denotes time derivative.  Equations (\ref{GPEs1}) and (\ref{GPEs2}) are the Euler-Lagrange equations of this Lagrangian density. The energy density $\cal E$ of a stationary state is the same as the Lagrangian density $\cal L$ setting the time-derivatives to zero.

To make the Lagrange variational analysis {\cite{vari}} analytically tractable, we use single-mode approximation (SMA), 
which implies that both the component wave functions have the same shape.
Indeed, in our numerical calculation we find that in many cases the component densities are very similar. 
Here we formulate a simple variational approximation in these cases. 
 In the present case, where the 
dimensionless wave functions are normalized to unity, SMA means
$|\phi_1|^2 = |\phi_2|^2$.

In a localized system, when $a_{12} = -\sqrt{a_1 a_2}$, which implies that $\sqrt{N_1}U_{12} = -\sqrt{N_2}\sqrt{U_1 U_2}$, 
the mean-field interaction energy  density without LHY and three-body interactions is 
\begin{eqnarray}
{\cal E}_{\rm int} 
            &=& \frac{1}{2} \left[\sqrt{N_1 U_1}n_1 - \sqrt{N_2 U_2}n_2\right]^2 ,\label{E/V_hom.}
 \end{eqnarray}  
{ where $n_i=|\phi_i|^2.$}
Minimizing  interaction energy density  (\ref{E/V_hom.}) with respect to densities  $n_1$ and $n_2$, we get
\begin{equation}
N_1 \sqrt{a_1}n_1 = {N_2}\sqrt{{a_2}} n_2. \label{den_ratio}
\end{equation}
Equation (\ref{den_ratio})  {implies} that if $N_1\sqrt{a_1} = N_2\sqrt{a_2}$, then 
 $|\phi_1|^2 = |\phi_2|^2$, which is  the condition for  SMA.
Hence if we  assume $a_{12} \approx -\sqrt{a_1 a_2}$ and choose $N_1\sqrt{a_1} = N_2\sqrt{a_2}$, SMA will be  a
reasonable approximation even after including the LHY and three-body interactions.  
Nevertheless, for the formation of a self-bound binary ball we require an attractive 
two-body energy density, which is possible for $a_{12} < -\sqrt{a_1a_2}$. To satisfy this condition, 
we take  $a_{12} = -(\delta a  +\sqrt{a_1a_2})$, where we take $\delta a $ to be small {and positive} 
so that the condition of SMA   remains approximately valid. 
Thus setting $|\phi|\equiv |\phi_i|, i=1,2,$ the Lagrangian density (\ref{Lagrangian}) becomes
  \begin{eqnarray}
{\cal L}& =  &    \frac{N}{2} \Big\{  {\mathrm i}  \big(  \phi   \phi^{*'} - \phi^*
 \phi ^{'}\big)+  \big| \nabla \phi\big|^2 
 \Big\}
-4\pi N_1N_2 \delta a |\phi|^4 \nonumber \\
& 
+ &\frac{256 \sqrt \pi}{15} \big(\sum_i N_i a_i\big)^{5/2} |\phi|^5+\frac{K_3}{6}N^3 |\phi|^6
 \label{Lagrangian2},
\end{eqnarray}
where $N=N_1+N_2.$
This Lagrangian density with the time-derivative terms and $K_3$ set equal to zero is the same as the stationary energy density given by  (1) of Ref. \cite{Cabrera}. The NLS equation in the SMA is the following Euler-Lagrange equation of Lagrangian density (\ref{Lagrangian2}):
\begin{eqnarray}
{\mathrm i} \frac{\partial \phi}{\partial t}  &= &\biggr[-\frac{ \tilde{\nabla}^2}{2} -\frac{8\pi N_1 N_2}{N} \delta a |\phi|^2 +\frac{{K}_3}{2} N^2 |\phi|^4   \nonumber \\
&+&\frac{128\sqrt \pi \Big(   \sum_i N_ia_i\Big) ^{5/2} }{3N} |\phi|^3 
 \biggr]\phi.\label{GPESMA} 
\end{eqnarray}
A Lagrange variational approximation to the SMA (\ref{GPESMA}) can be performed with the following 
 variational {\em ansatz} {\cite{vari}} 
\begin{eqnarray}
\phi ={\pi ^{-3/4} w^{-3/2}} {\exp \Big({-\frac{r^2}{2 w^2}+\mathrm{i} \kappa r^2}\Big)},\label{gaussian_ansatz}
\end{eqnarray}
where $w$ is the width and $\kappa$ the chirp.
{
 Although we are looking for a real density, the wave function is complex and has a phase. The chirp term is a simple way of introducing the phase. This term is necessary to get the time-dependent dynamics, viz. Eq. (\ref{EL_eq}). For a stationary solution we can set this term to zero.}
 Using this {\em ansatz}, the Lagrangian density   (\ref{Lagrangian2}) can be integrated over all space to yield the   Lagrangian functional  
\begin{eqnarray}\label{lagr}
L &\equiv & \int {\cal L}d{\bf r} ={3}\kappa ^2 w^2N+\frac{3}{2} w^2 \kappa 'N+\frac{3 N}{4 w^2}
- \frac{8\pi N_1 N_2 \delta a}{4 \sqrt{2} \pi ^{3/2} w^3} 
\nonumber \\
&
+ &\frac{512 \sqrt{{2}} \left(\sum_i N_ia_i\right){}^{5/2}}{75\sqrt 5  \pi ^{7/4} w^{9/2}} 
+\frac{K_3 N^3}{18 \sqrt{3} \pi ^3 w^6},
\end{eqnarray}
where for this study of a self-trapped binary ball, we have removed the contribution of the 
trapping potential $V$.
{
If $w$ is the characteristic size of the condensate, the kinetic energy, interaction energy, LHY interaction energy, and three-body interaction energy  scale as $w^{-2}, -w^{-3},
w^{-9/2}$, and $w^{-6}$, respectively. This energy scaling suggests that a minimum in energy  (local or global) can occur for a finite size $w$. This vindicates the use of variational ansatz (\ref{gaussian_ansatz}) having a characteristic width $w$ like a Gaussian function, specially for the tightly bound balls. Nevertheless the homogeneous equation (\ref{GPESMA}) (setting the non-linear terms to zero) is a plane-wave equation.
Using the definition of Fourier transform the Gaussian ansatz (\ref{gaussian_ansatz}) can be considered as the  superposition of an infinite number of plane waves. This justifies the use of a Gaussian ansatz in Eq.  (\ref{GPESMA}).}

In the absence of the last two terms in Eq. (\ref{lagr}) with LHY and three-body energy contributions, 
the Lagrangian (or the energy) of a stationary state obtained by setting $\kappa=0$ in  (\ref{lagr}) tends to $-\infty$ as $w\to 0$ signaling a collapse instability. However, in the presence of any or both of these terms the energy at the center ($w=0$) becomes infinitely large  and hence a collapse is avoided. 
The Euler-Lagrange equations of the Lagrangian (\ref{lagr}) for variable $\nu  \equiv \kappa, w$, 
\begin{equation}
\frac{\partial }{\partial t}\frac{\partial L}{\partial \nu '}-\frac{\partial L}{\partial \nu }=0,
\end{equation}
lead to  
\begin{eqnarray}
 w'' &=  &\frac{ 1}{ w^3}- \frac{ 4\pi N_1N_2 \delta a}{\sqrt{2} \pi ^{3/2} w^4 N}  
 +
\frac{512 \sqrt 2 \left(\sum_i N_ia_i  \right){}^{5/2}}{25 \sqrt 5 N\pi ^{7/4} w^{11/2}} + \frac{2 K_3 N{}^2}{9 \sqrt{3} \pi ^3 w^7}, \label{EL_eq}
\end{eqnarray}
which describes the variation of the width of the condensate with time.  The width of a stationary binary ball is obtained by setting $w''=0 $ in  (\ref{EL_eq}).

\begin{figure}[t]
\begin{center}
\includegraphics[trim = 0.225cm -0.1cm 0cm 0cm, clip,width=.45\linewidth,clip]{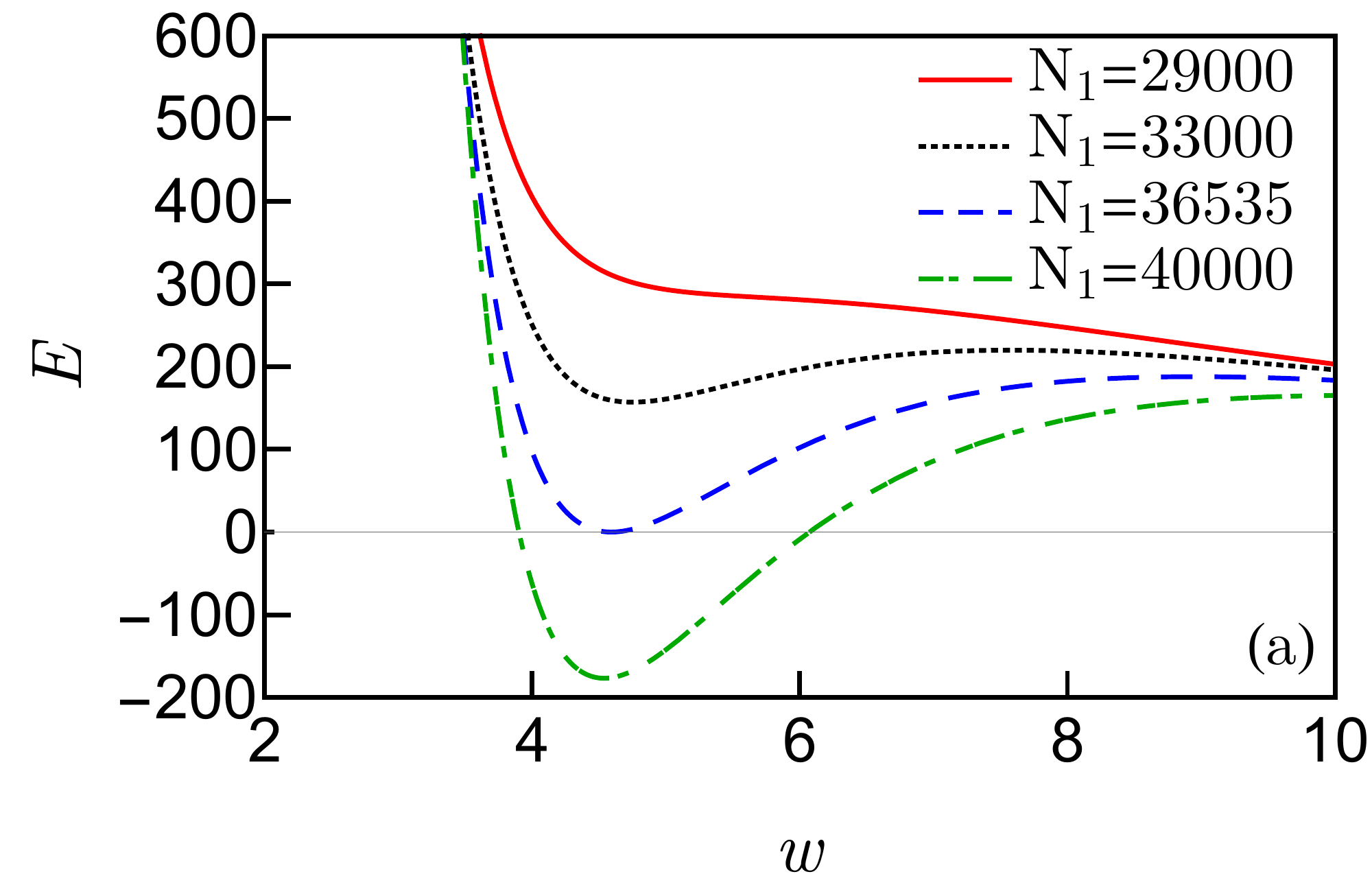}
\includegraphics[trim = 0cm 0cm 0cm -0.1cm, clip,width=.54\linewidth,clip]{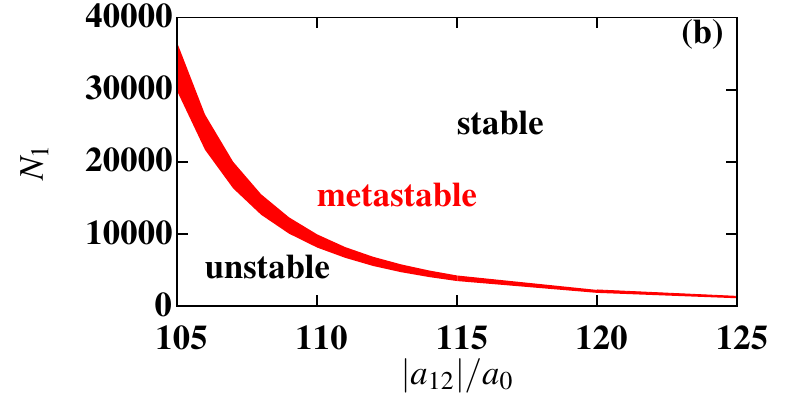}
\caption{(a) Variational energy of a binary  stationary Rb$^{87}$ ball with 
  {$a_{12} = -105 a_0$ and 
 $N_1 = 29000,~33000,~36535,~40000$}  as a function of {the} width of the condensate. 
(b) Phase plot  in $N_1-|a_{12}|$ plane showing the formation of a stable and 
meta-stable binary ball. Other parameters used are
 $a_1 = 100.4 a_0$, $a_2 = 95 a_0$, $K_3 = 1.8\times 10^{-41}$ m$^6$/s, and 
$N_2 = N_1\sqrt{a_1/a_2}$. The plotted quantities in this and following figures are dimensionless.}  
\label{fig1}
\end{center}
\end{figure}

\section{Numerical Result}

\label{III}

We use split time-step Fourier pseudo-spectral method to solve the coupled NLS 
equations (\ref{GPEs1})-(\ref{GPEs2}) numerically \cite{Bao}.
We  also cross-checked our results with split time-step Crank-Nicolson method \cite{Muruganandam}.  
The minimum-energy  ground-state solution for the binary ball 
is obtained by evolving the trial wave functions, chosen to be Gaussian, in imaginary
time $\tau = i t$ using  (\ref{GPEs1})-(\ref{GPEs2}) as is proposed in Refs. \cite{Muruganandam}. 
{
In numerical calculation we employ  periodic boundary condition, and the size of numerical domain is  taken sufficiently larger than the typical size of the self-trapped solution to make the boundary effects to be negligible.}
The numerical results  of the model for stationary binary balls presented in Sec. \ref{IIIA} are
obtained using spherical coordinate $r$. 
The spatial and time steps used to solve the NLS equations in imaginary time are 
$r=0.0025$ and $t=3.125\times10^{-6}$. We consider a binary ball consisting of $|F = 1, m_F= +1\rangle$ and $| F = 2, m_F= -1\rangle$ hyperfine states of 
$^{87}$Rb, which has been realized experimentally \cite{Tojo}. We term these as  components 1 and
 2, respectively, in the rest of the paper. The intra-species $s$-wave scattering lengths of  
the two components are $a_1=100.4a_0$ and $a_2=95a_0$ for components 1  and 2, respectively \cite{Tojo}, with $a_0$ the Bohr radius.
The inter-species scattering length for the system  $a_{12}$ is tunable with a magnetic Feshbach resonance \cite{fesh},
and can thus be used to realize the quantum ball. The collision dynamics of the moving binary balls is studied 
by real-time propagation in Cartesian coordinates
with space and time steps $0.04$  and $0.0004$, respectively, using the initial 
states obtained by imaginary-time propagation.

\subsection{Stationary quantum balls}

\label{IIIA}

We consider the formation of a binary stationary quantum ball of $^{87}$Rb atoms 
in two hyperfine states with {$N_1=30000$ and 50000} atoms in the first state. 
The number of atoms in the second state is taken as  $N_2 = N_1\sqrt{a_1/a_2}$ (chosen to make SMA a good approximation).
The inter-species scattering length is taken as {$a_{12} = -105 a_0$.} The  
scaling length used to write dimensionless NLS equation is   $l_0=1$ $\mu$m; {the corresponding scaling time is $ml_0^2/\hbar =1.37$ ms.}
Actually, there is no estimate of $K_3$, the three-body interaction strength and specially, its real part, 
which helps in the formation of the binary ball. However, an estimate for the three-body loss rate $-$ the imaginary part of $K_3$ $-$  of $^{87}$Rb exists:  
$K_3 = 1.8\times 10^{-41}$ m$^6$/s \cite{Tojo,tojo2}. Because of unitarity constraints the real 
part of $K_3$ should have a magnitude of the same order.  In this study we take the real 
and imaginary parts of $K_3$ to be identical, e.g., $K_3=  1.8\times 10^{-41}(1-i)$ m$^6$/s. 
Nevertheless, in the study of the stationary quantum ball we set the imaginary part of $K_3$ to zero, 
and later we find in the study of dynamics  that the effect of the imaginary part is
{ negligible.  
The} numerical solution for the ground-state binary ball is obtained by imaginary-time propagation and 
 the variational solution is obtained 
from a solution of  (\ref{EL_eq}) with $w'' = 0$.

\begin{figure}[t]
\begin{center}
\includegraphics[trim = 0cm 0cm 0cm 0cm, clip,width=0.49\linewidth,clip]{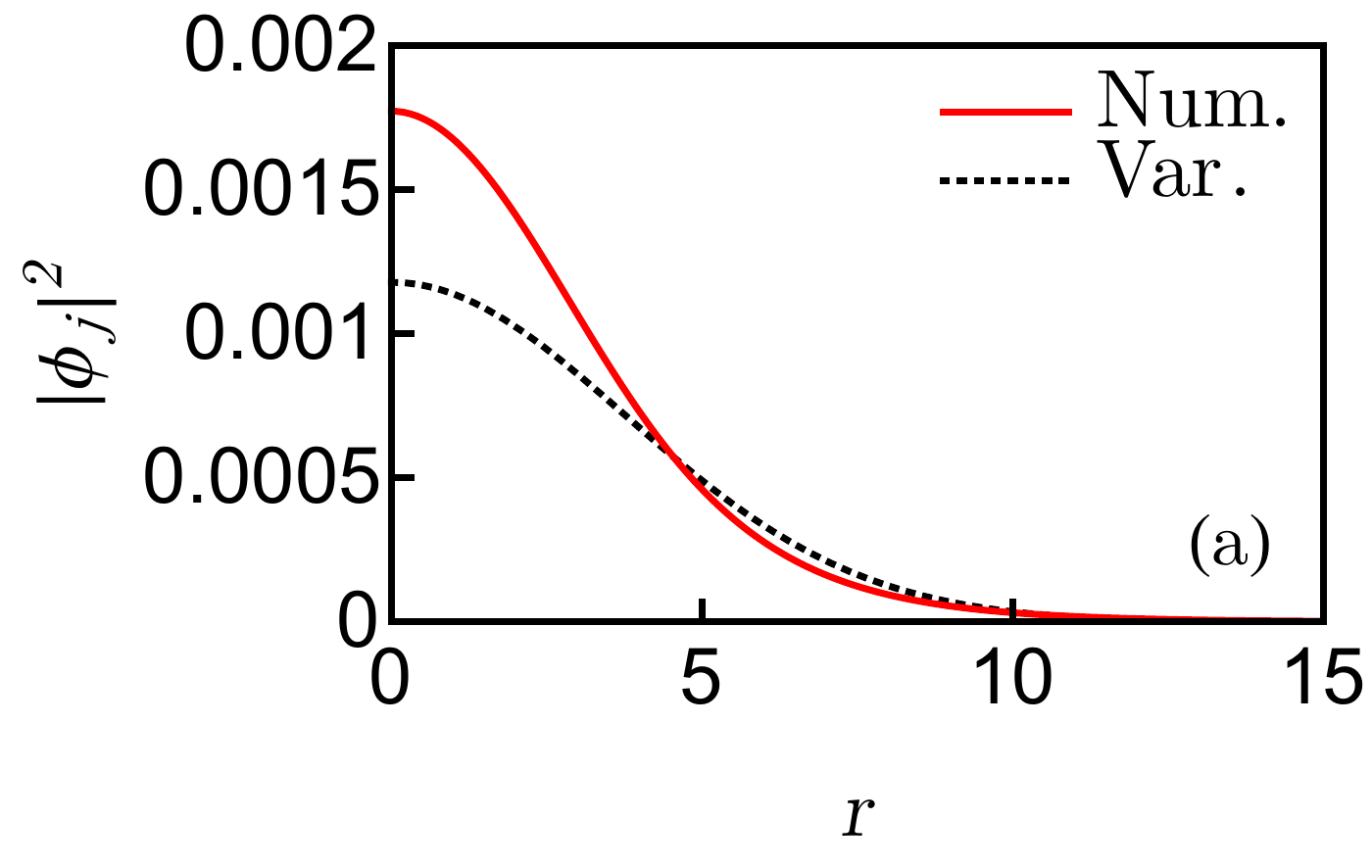} 
\includegraphics[trim = 0cm 0cm 0cm 0cm, clip,width=0.49\linewidth,clip]{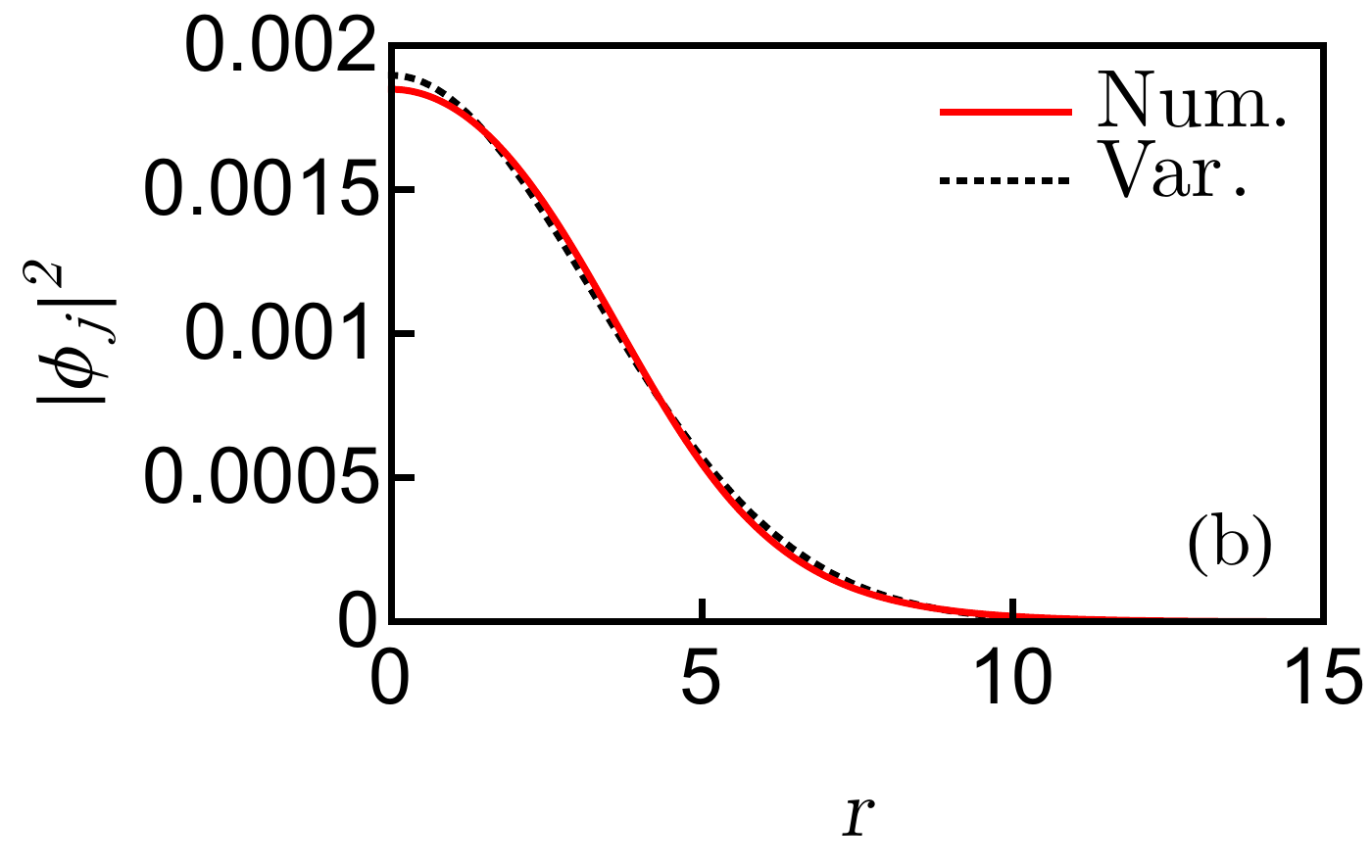}
\includegraphics[trim = 0cm 0cm 0cm 0cm, clip,width=0.49\linewidth,clip]{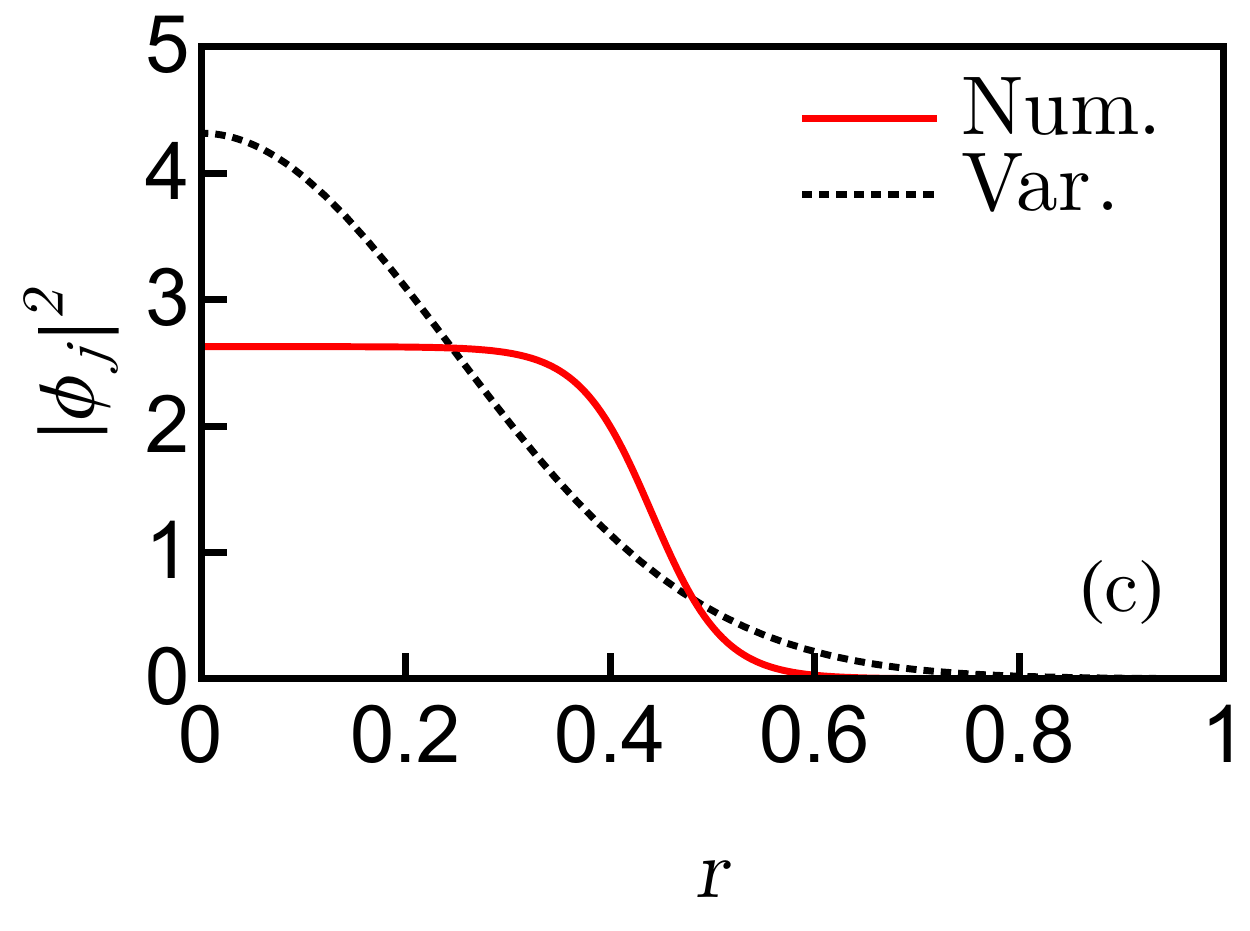}
\includegraphics[trim = 0cm 0cm 0cm 0cm, clip,width=0.49\linewidth,clip]{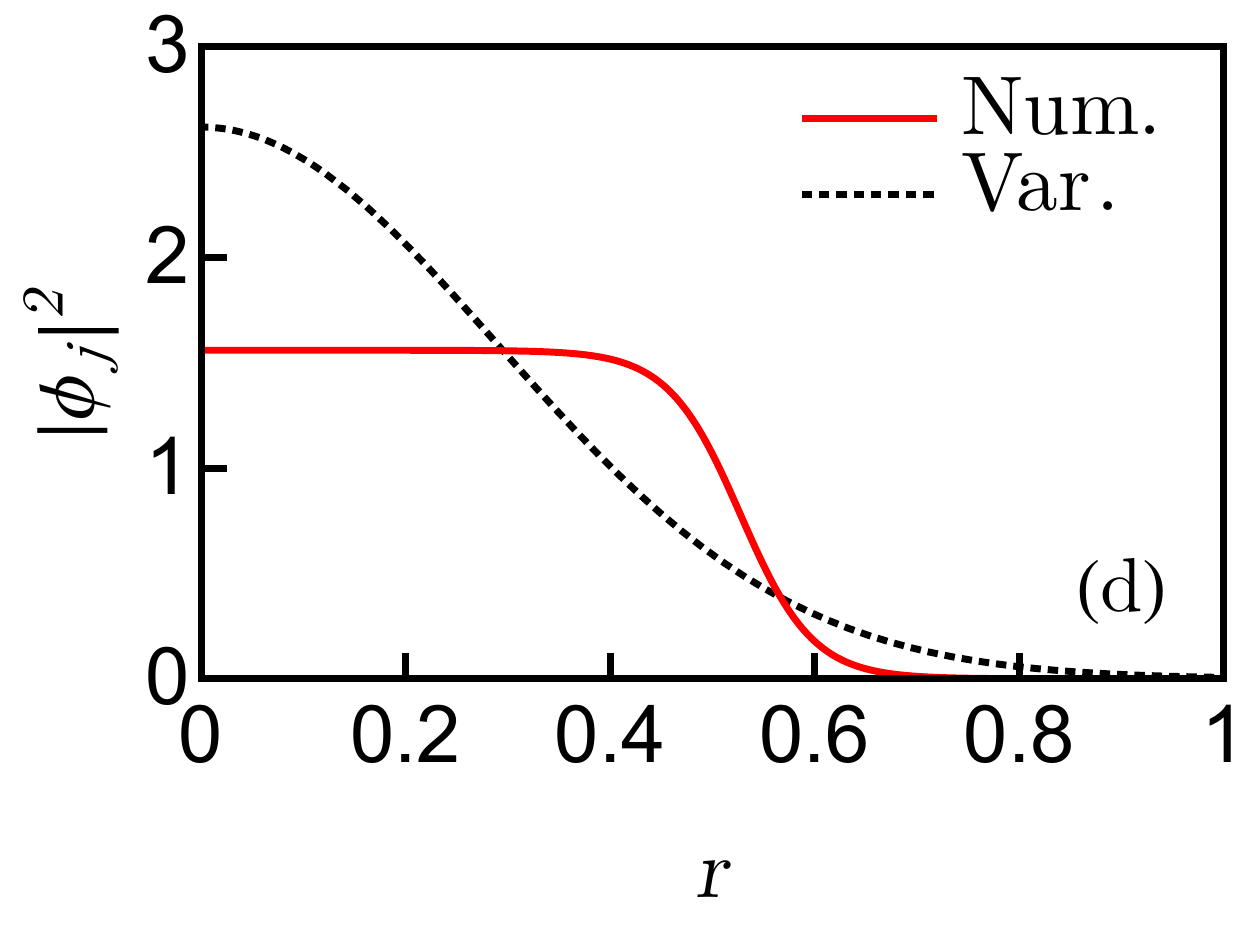}
\caption{ {Numerical {(Num.)}  and variational {(Var.)} density  of a binary   $^{87}$Rb ball  with 
$a_1 = 100.4 a_0$, $a_2 = 95 a_0$, $a_{12} = -105 a_0$, $K_3 = 1.8\times 10^{-41}$ m$^6$/s, and
 $N_2 = N_1\sqrt{a_1/a_2}$  for 
 (a) $N_1 = 30000$, 
and (b) $N_1 = 50000$.    (c) and (d) display  the densities for
the same parameters as in  (a) and (b), respectively, but with the LHY interaction switched off. }}
\label{fig2}
\end{center}
\end{figure}

The variational results confirm the existence of energetically meta-stable as well as stable  balls. To illustrate the
distinction between a meta-stable and a stable  ball, the variational energy $E$ of  binary balls
with aforementioned scattering lengths as a function of the variational width $w$ are plotted in figure \ref{fig1}(a).
Incidentally, the variational energy $E$ is the Lagrangian (\ref{lagr}) with $\kappa=0$.
In figure \ref{fig1}(a), a meta-stable  ball corresponds to a curve with a local minimum in the energy, 
whereas   a stable ball corresponds to the curve with a global minimum. These two cases are shown in figure \ref{fig1}(a) 
for {$N_1 = 33000$ and $N_1 = 40000$,} respectively. An unstable ball corresponds to a curve with 
no minimum, viz. {$N_1=29000$} in figure \ref{fig1}(a). 
 The variational phase plot of the system in 
the $N_1-|a_{12}|$ plane, while
keeping intra-species scattering lengths and $K_3$ fixed, is shown in figure \ref{fig1}(b), which clearly shows the
regions of energetically stable and meta-stable balls. In the following we will study only the stable binary balls.

\begin{figure}[t]
\begin{center}
\includegraphics[trim = 0cm 0cm 0cm 0.15cm, clip,width=0.49\linewidth,clip]{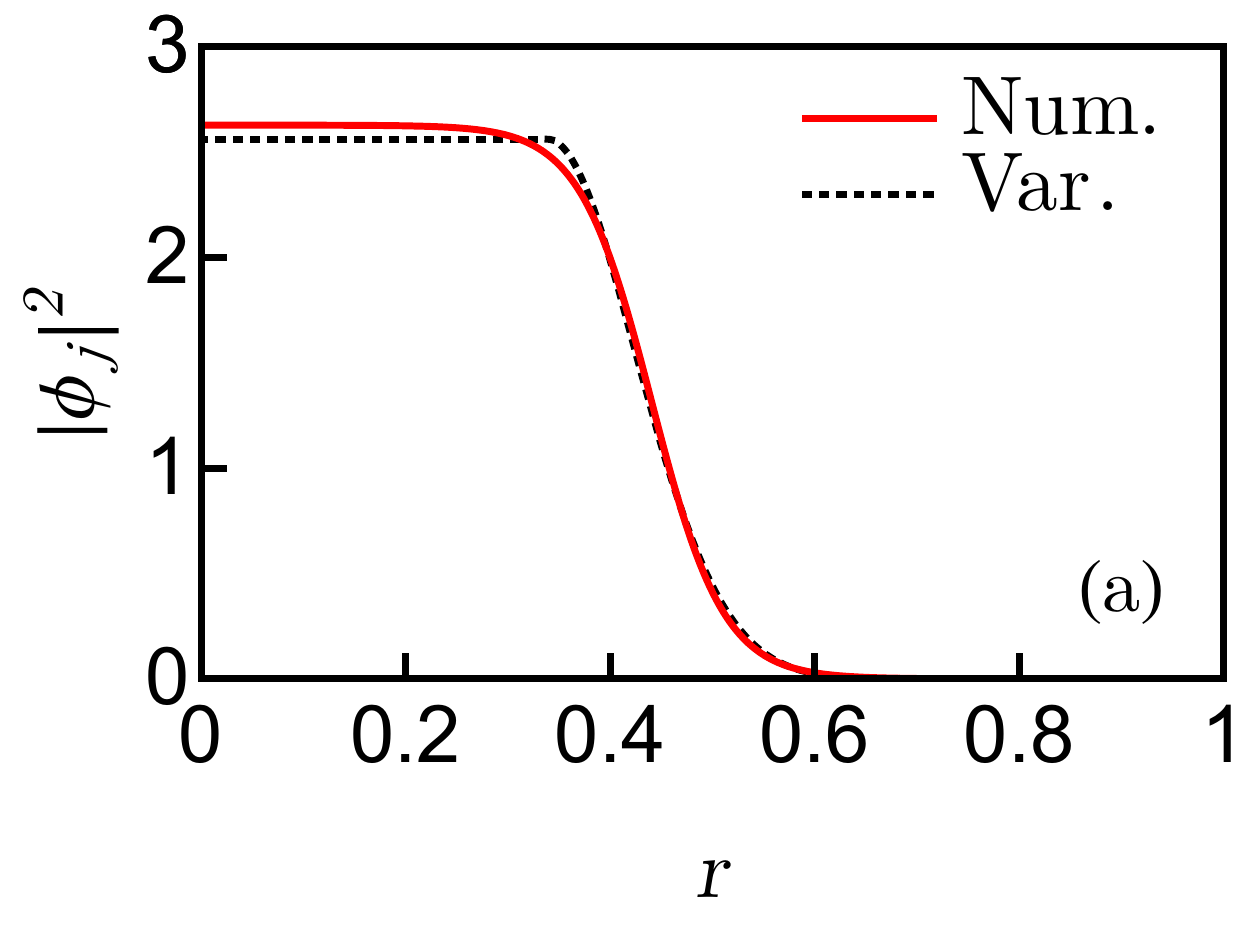}
\includegraphics[trim = 0cm 0cm 0cm 0.15cm, clip,width=0.49\linewidth,clip]{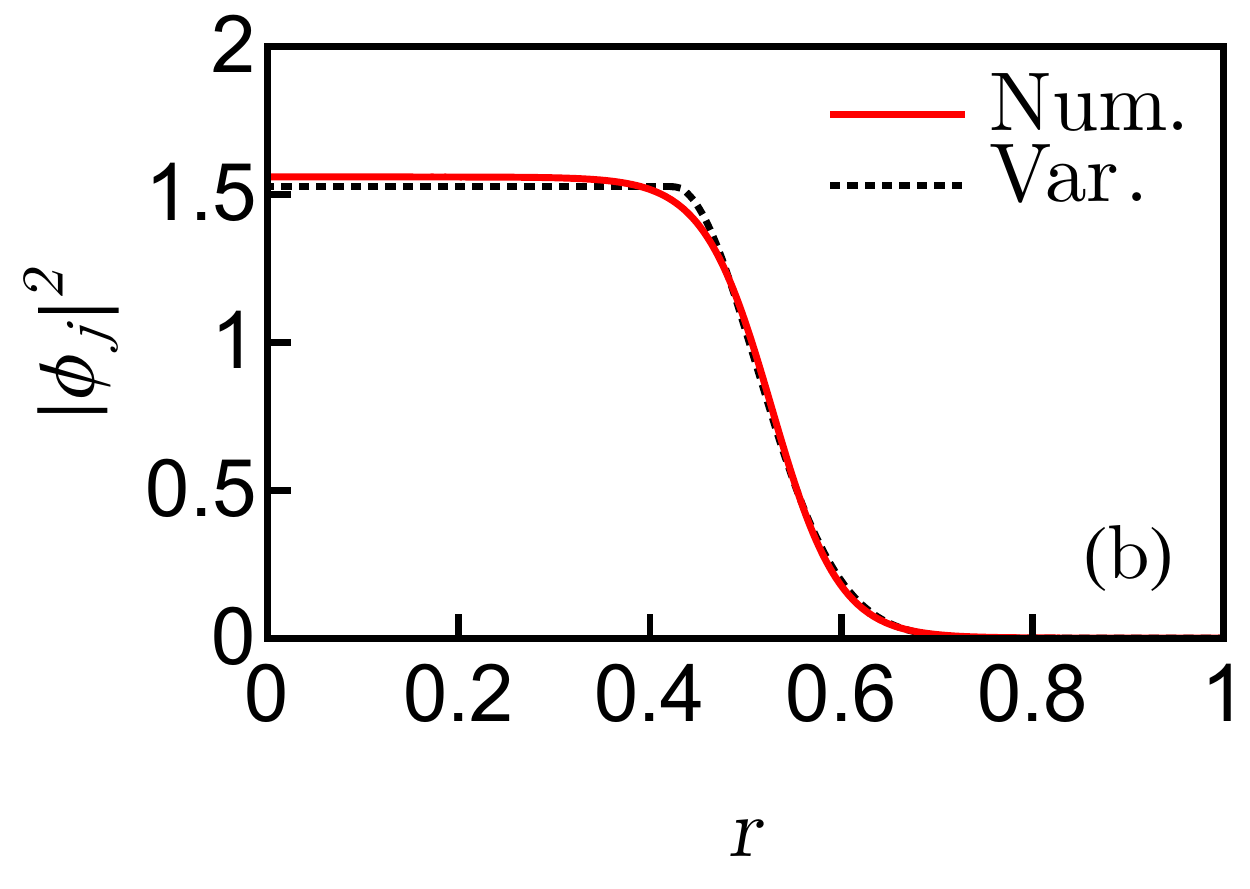}
\caption{{Numerical {(Num.)}  and variational {(Var.)} density  of a  binary   $^{87}$Rb ball  for (a)     $N_1 = 30000$ and 
(b) $N_1 = 50000$. All other parameters are the same as in figure \ref{fig2}(c) and (d), respectively. 
Variational results are calculated with  {\em ansatz}  (\ref{Piece_wise}).}}
\label{fig3}
\end{center}
\end{figure}

\begin{figure}[t]

\begin{center}
\includegraphics[trim = 0cm 0cm 0cm 0cm, clip,width=0.49\linewidth,clip]{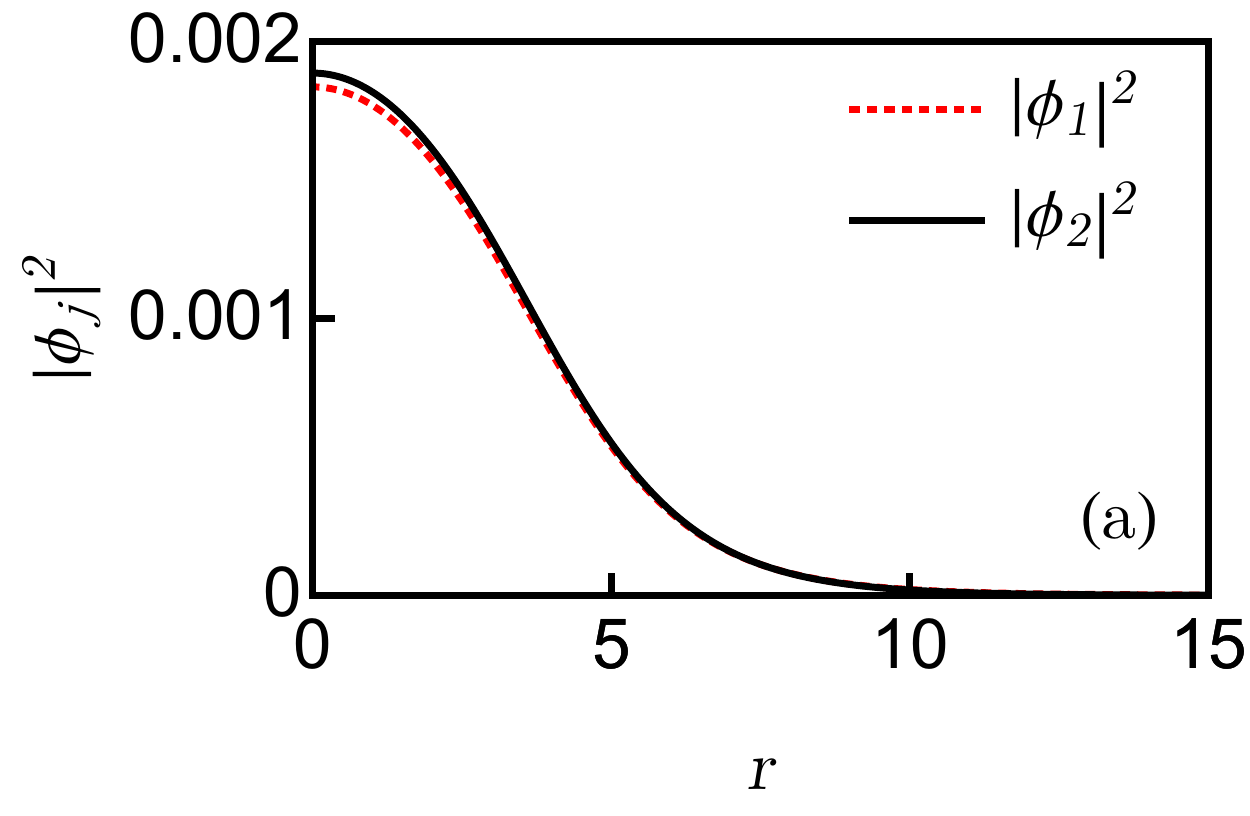}
\includegraphics[trim = 0cm 0cm 0cm 0cm, clip,width=0.49\linewidth,clip]{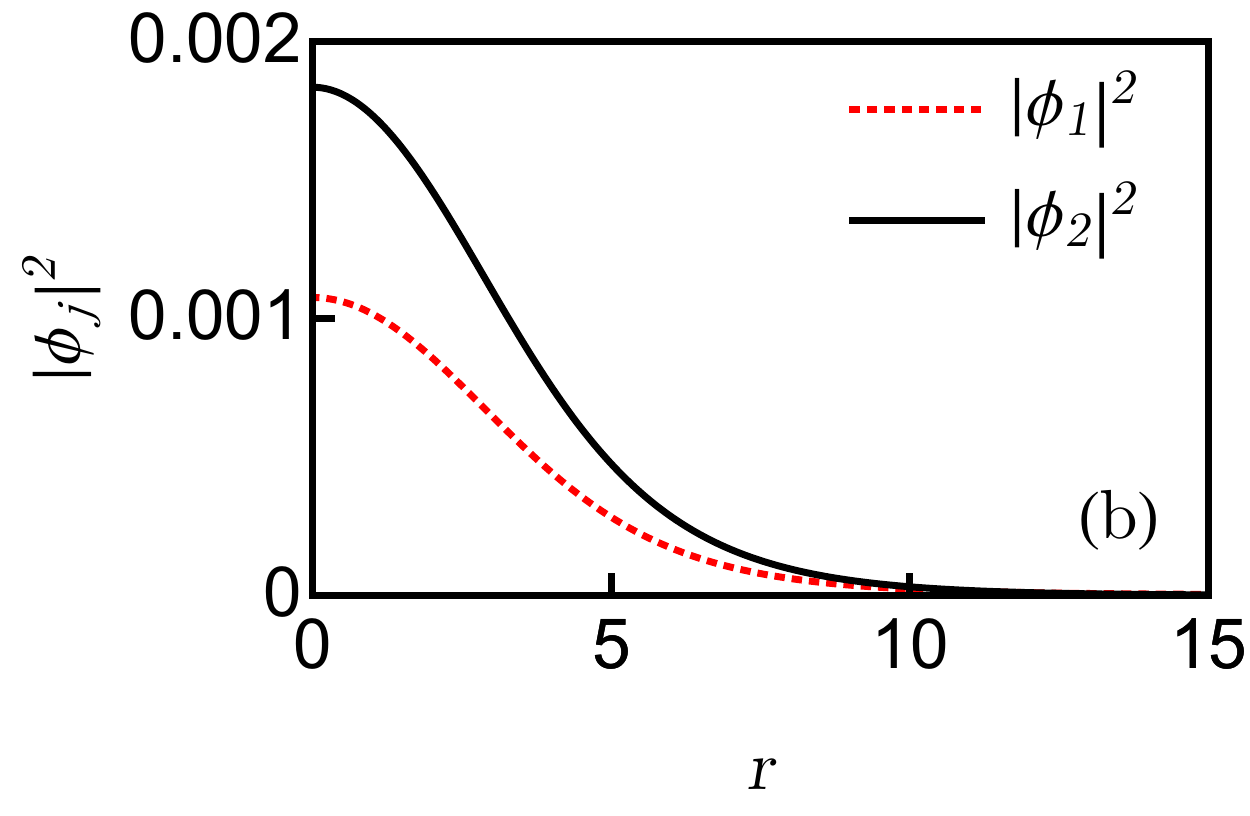}
\caption{{Numerical density of a binary   $^{87}$Rb ball  with (a)
$N_1 = N_2 = 50000$, and  (b) $N_1 = 50000$, $N_2 = 30000$. 
{
Other parameters used are
 $a_1 = 100.4 a_0$, $a_2 = 95 a_0, a_{12}=-105 a_0,$ and $K_3 = 1.8\times 10^{-41}$ m$^6$/s.}}
}
\label{fig4}
\end{center}
\end{figure}

In figures \ref{fig2}(a)-(d), we display the numerical and variational  common densities $|\phi_j|^2$  {with $j=1,2$} of the two components 
of the binary ball for different parameters.  
The variational result with SMA (\ref{EL_eq}) is in good agreement with the numerical solution of the  NLS equations  (\ref{GPEs1})-(\ref{GPEs2})
with LHY  and three-body interactions as is shown in  figures \ref{fig2}(a)-(b).
In figures \ref{fig2}(c)-(d), density profiles of the binary ball after switching off the LHY interaction
are also shown. The variational densities  in this case are poor approximations to the numerical ones. 
This is due to the fact that in the absence of the LHY interaction, the repulsive force needed for the stabilization of the
quantum ball is provided by the kinetic energy and the  three-body interaction. If the characteristic width of the ball 
$w\ll1$, then the three-body interaction energy, which varies as $w^{-6}$, is much larger than the kinetic energy, which varies as $w^{-2}$, 
as is evident from their respective contributions in the Lagrangian (\ref{lagr}). Therefore to model the densities,  
in the absence of the  LHY interaction for small $K_3$, one can neglect the kinetic energy in comparison
to mean-field two- and three-body interaction energies, except near the surface of the ball, 
in the NLS equations (\ref{GPEs1})-(\ref{GPEs2}). This will lead to constant density profiles 
for the components in the bulk of the binary ball. The kinetic energy nevertheless will contribute to the energy near the surface of the  ball 
where the wave function would decay  to zero over some length. This motivates the use of the following piece-wise continuous variational
{\em ansatz} for the component densities 
\begin{equation}
\begin{array}{cc}
 \phi_j (r)=\Big\{ & 
\begin{array}{cc}
 b, & 0<r\leq R_{\mathrm{in}} \\
 b \exp  \Big[ {-\frac{ \left(r-R_{\mathrm{in}}\right){}^2}{2 l^2}}\Big], & R_{\mathrm{in}}\leq r< \infty,\label{Piece_wise}  \\
\end{array}
 \\
\end{array}
\end{equation}
where $l$, $b$, $R_{\rm in}$  are the variational parameters; $R_{\rm in}$ is the radius of the 
core within which kinetic energy can be  neglected, and $l$ is the characteristic length over which the wave function decays to
zero from its constant value of $b$. Normalization will fix
one of these, say $b$, leaving us with two independent variational parameters. Minimizing the energy
of the binary system numerically, one can determine the variational parameters. The variational densities 
using {\em ansatz}  (\ref{Piece_wise}) are now in good agreement with the numerical results as
  illustrated in figures {\ref{fig3}}(a)-(b).

\begin{figure}[t]
\begin{center}
\includegraphics[trim = 0cm 0cm 0cm 0cm, clip,width=0.49\linewidth,clip]{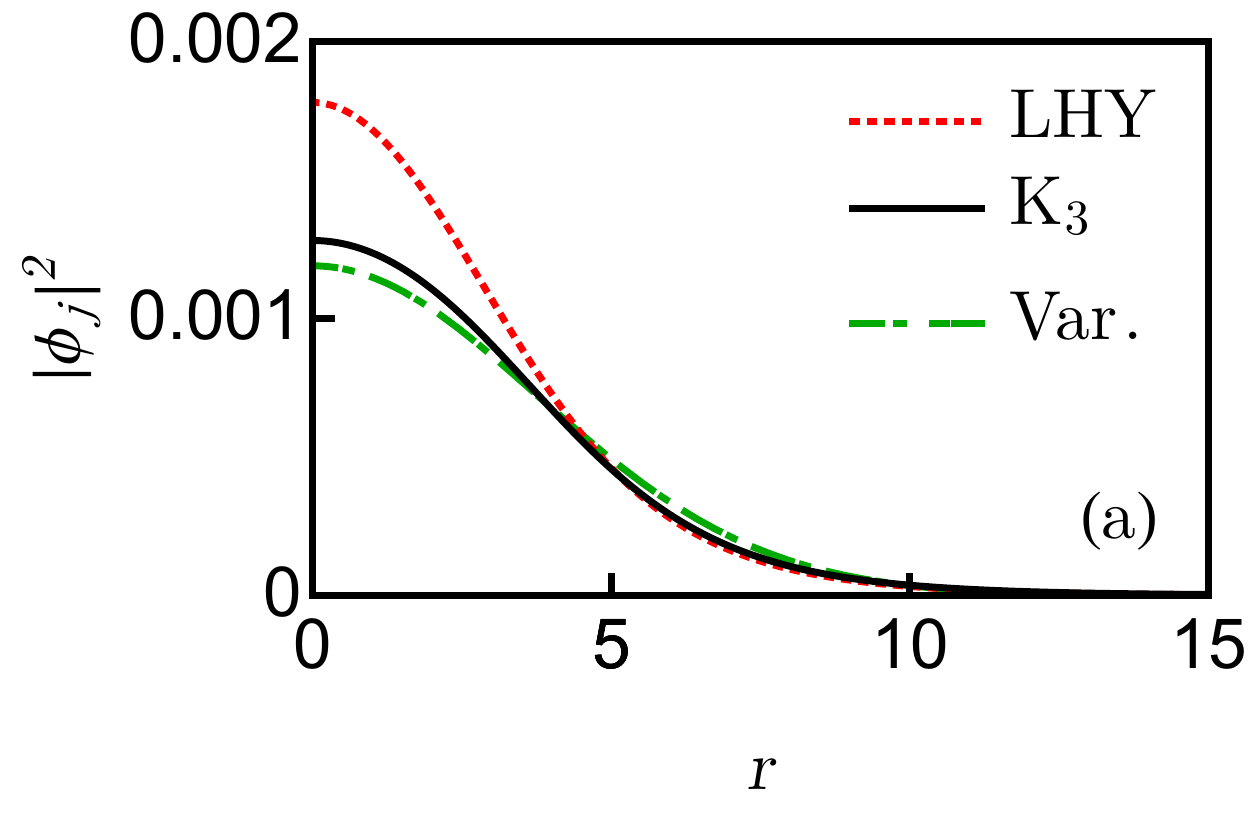}
\includegraphics[trim = 0cm 0cm 0cm 0cm, clip,width=0.49\linewidth,clip]{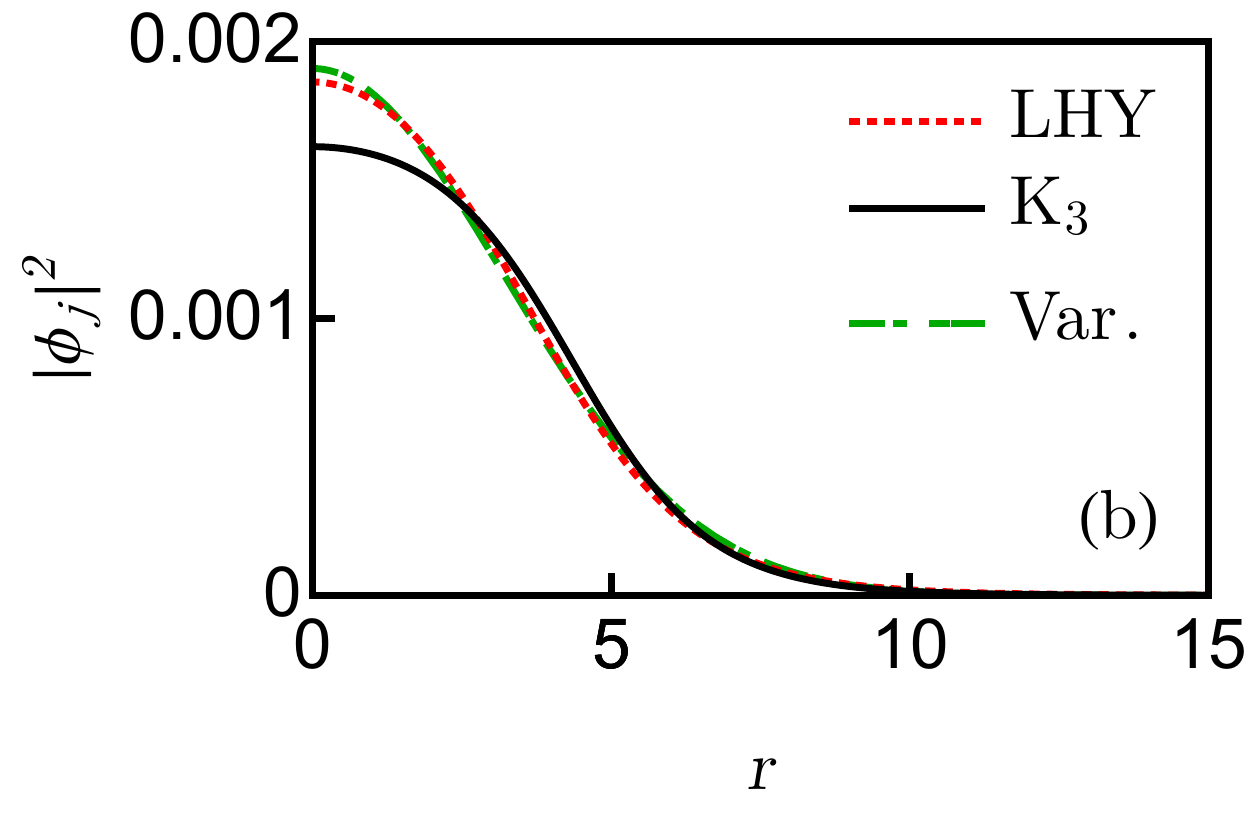}
\caption{ { Numerical density  of a binary  $^{87}$Rb ball  with (a)
$N_1 = 30000$,   and (b) $N_1=50000,$  with only LHY interaction (LHY) and
with only a tuned three-body interaction ($K_3$). The variational density   {(Var.)}, which is
the 
same in both   cases, is also shown.   The tuned
$K_3$, calculated from  (\ref{K3_estimate}), is $2.95\times 10^{-38}$m$^6$/s and
$=1.8\times 10^{-38}$m$^6$/s for (a) and (b), respectively. {Other parameters used are
 $a_1 = 100.4 a_0$, $a_2 = 95 a_0$, $a_{12}=-105a_0$, and 
$N_2 = N_1\sqrt{a_1/a_2}$}.}
}
\label{fig5}
\end{center}
\end{figure}

{\em Break-down of SMA:} If the condition for the validity of SMA, i.e.  
the condition $N_1\sqrt{a_1} = N_2\sqrt{a_2}$,
 is not satisfied then the variational analysis 
discussed above  is no longer applicable. For example, considering {$N_1 = N_2 = 50000$} and keeping the interaction parameters same
as in figure \ref{fig2} with both three-body interaction and LHY interaction  switched on, the numerical density 
of the binary  ball 
 is shown in figure \ref{fig4}(a). In this case, the densities of the two components are
no longer overlapping. The break-down of SMA is much more pronounced in the case with {$N_1 = 50000$, $N_2 = 30000$}
as  illustrated in  figure \ref{fig4}(b). This is because with the parameters of figure \ref{fig4}(b) 
the violation of the SMA condition $N_1\sqrt{a_1} = N_2\sqrt{a_2}$ is stronger than with  the parameters of figure \ref{fig4}(a).

{ As mentioned in   Sec. \ref{Sec-I}, both the LHY as well 
three-body interactions, acting independently or jointly, can stabilize the binary ball. It implies that if three-body interaction coefficient $K_3$ is assumed
to be tunable {\cite{tune}}, then by switching off the LHY interaction, one can tune $K_3$ to obtain a binary ball  of
 similar shape and size as one would obtain with only the LHY interaction.
If $w_0$ is the variational width of the binary ball with only LHY interaction, then using  (\ref{EL_eq}),
the $K_3$ needed to obtain the same width with only three-body interaction is
\begin{equation}
K_3=\frac{2304 \sqrt 6\big(\sum_i N_i a_i \big)^{5/2}}{25 \sqrt 5  N^3}  w_0^{3/2}\pi^{5/4}.
\label{K3_estimate}
\end{equation} 
For {
$N_1 = 30000$, $N_2 = N_1\sqrt{a_1/a_2}$, 
$a_1 = 100.4 a_0$, $a_2 = 95 a_0$, $a_{12} = -105 a_0$, the variational width obtained by using Gaussian {\em ansatz}, namely
 (\ref{gaussian_ansatz}), with only LHY interaction is $w_0 = 5.32$. Using this in  (\ref{K3_estimate}),
the $K_3$   needed to obtain the same variational width with only three-body interaction  
is $2.95\times 10^{-38}$m$^6$/s. Similar calculation for $N_1 = 50000$, $N_2 = N_1\sqrt{a_1/a_2}$ gives $K_3 = 1.8\times 10^{-38}$m$^6$/s.}
The numerical and variational results for density  with only LHY interaction and with only 
a tuned three-body interaction
are shown in figures \ref{fig5}(a)-(b), which look quite similar.

\begin{figure}[t]
\begin{center}
\includegraphics[trim = 0cm 0cm 0cm 0cm, clip,width=0.49\linewidth,clip]{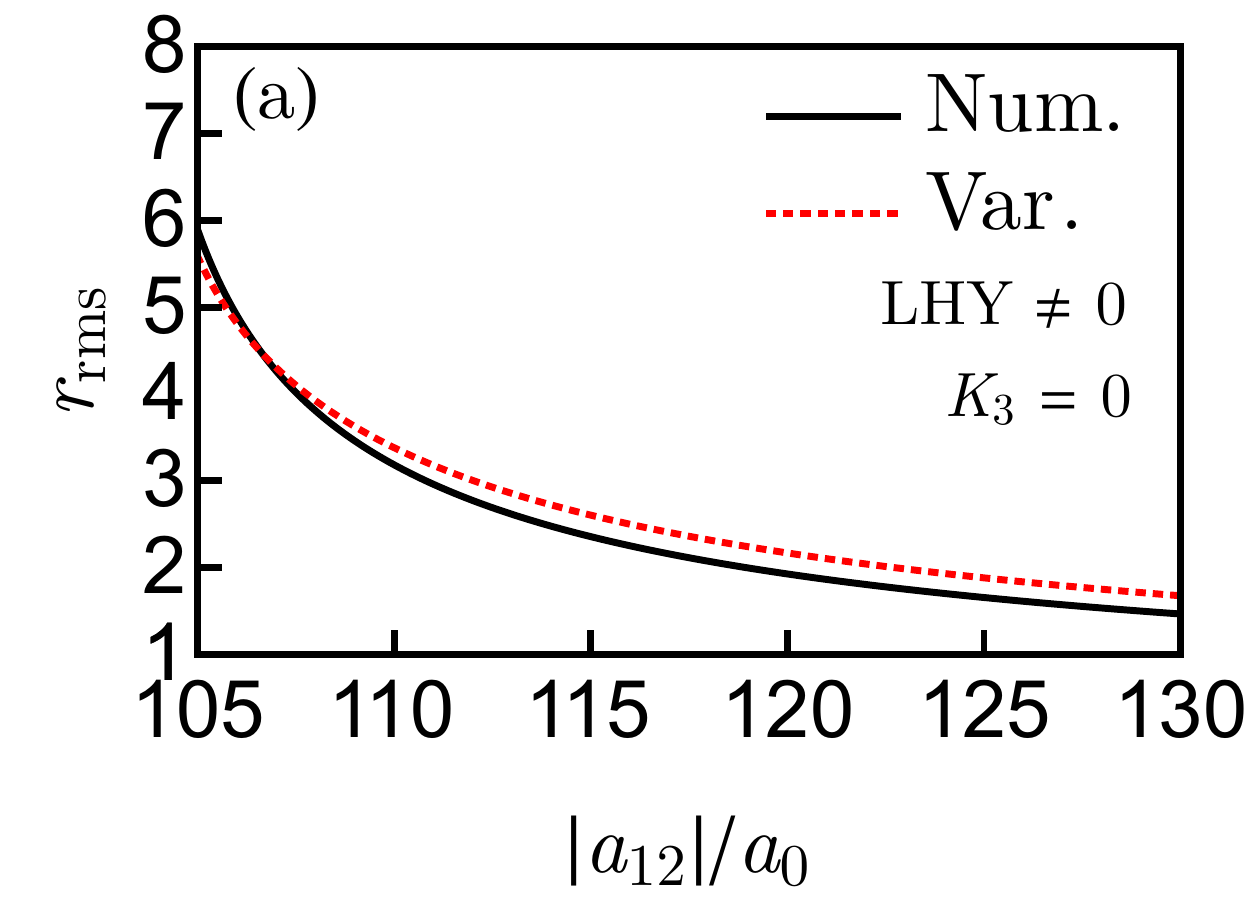}
\includegraphics[trim = 0cm 0cm 0cm 0cm, clip,width=0.49\linewidth,clip]{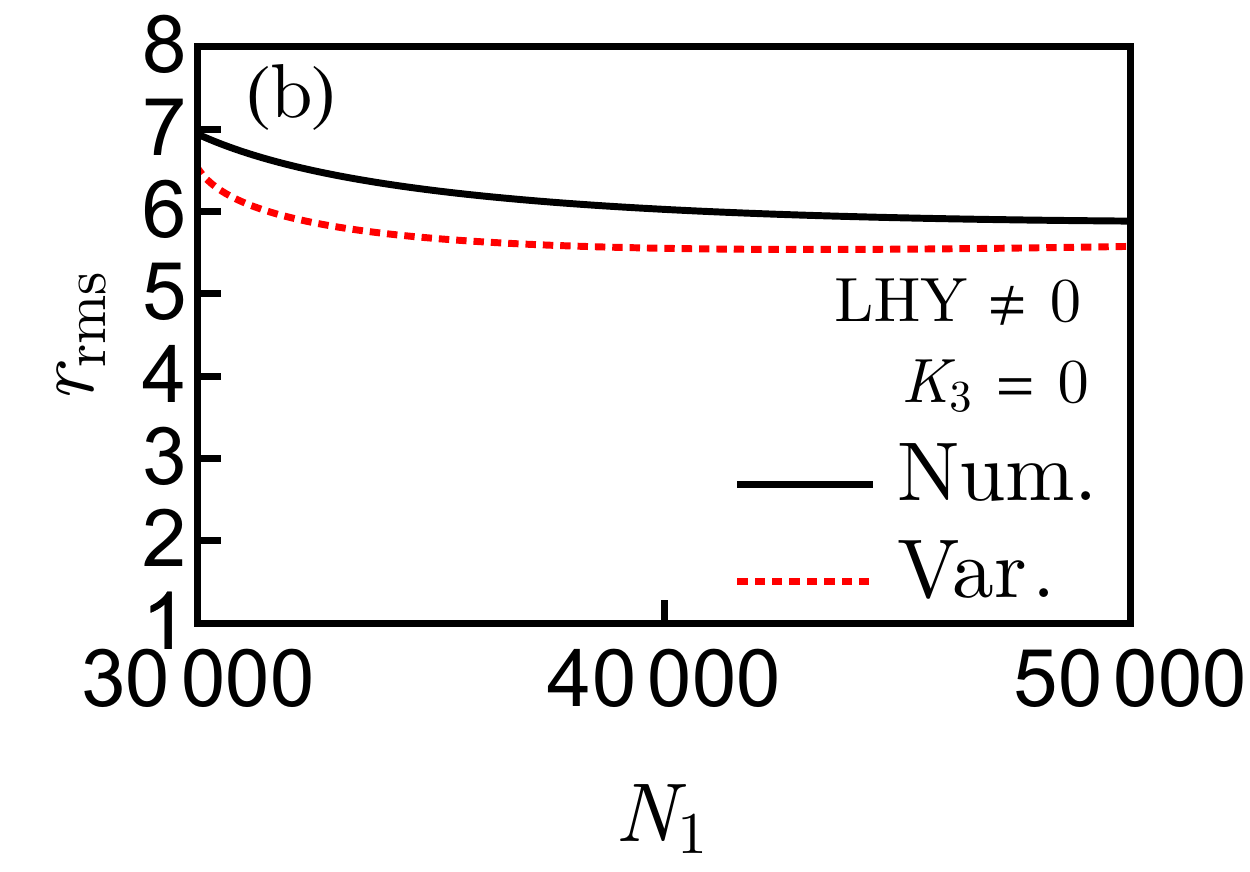}
\includegraphics[trim = 0cm 0cm 0cm 0cm, clip,width=0.49\linewidth,clip]{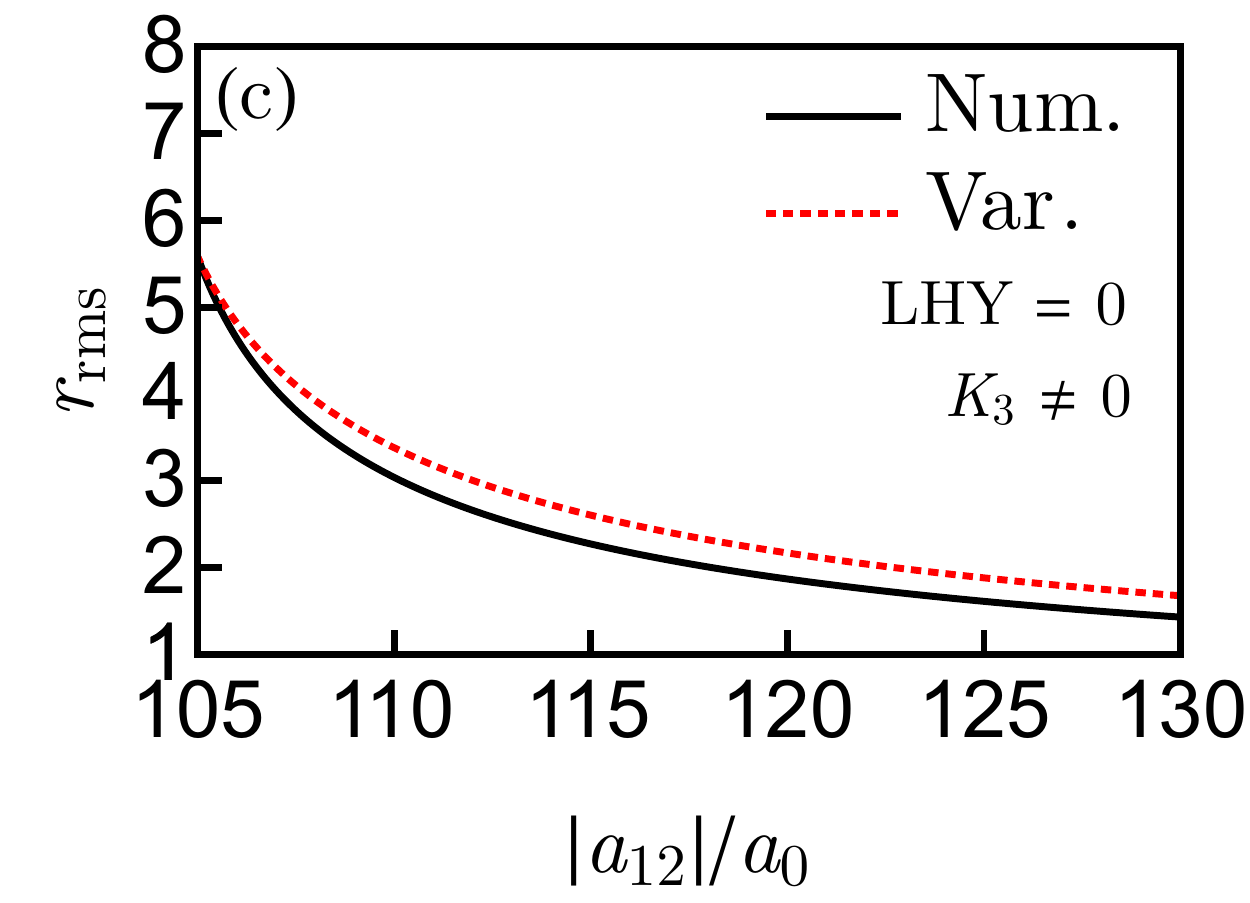}
\includegraphics[trim = 0cm 0cm 0cm 0cm, clip,width=0.49\linewidth,clip]{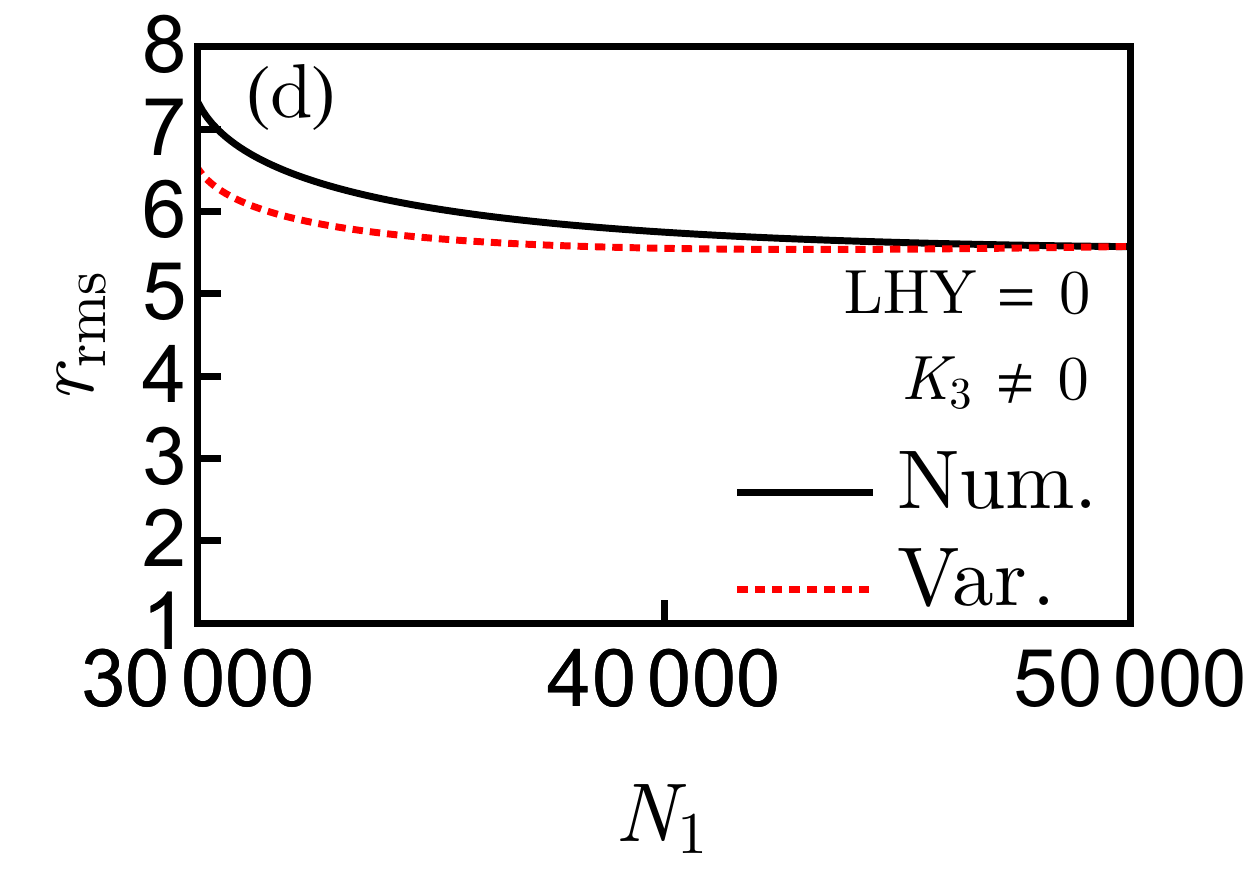}
\caption{ {(a) Numerical (Num.) and variational (Var.)  rms sizes of a 
binary   $^{87}$Rb ball with
$N_1 = 50000$,   $K_3 = 0$,  
as a function of $a_{12}$; (b) the same as a function of $N_1$ with  
  $a_{12} = -105a_0$; (c) the same with zero LHY interaction and 
$N_1 = 50000$,  $K_3 \ne 0$, tuned using  (\ref{K3_estimate}), 
as a function of $a_{12}$; (d)  the same with zero LHY interaction  as a function of $N_1$ with  
$a_{12} = -105a_0$, $K_3\ne0$, tuned using  (\ref{K3_estimate}).
 Other parameters used are
 $a_1 = 100.4 a_0$, $a_2 = 95 a_0$,   and 
$N_2 = N_1\sqrt{a_1/a_2}$.
}
}
\label{fig20}
\end{center}
\end{figure}

We also studied the variation of the root-mean-square (rms) sizes of the stationary quantum balls  for a fixed number of atoms $N_1$ 
as a function of  the  inter-species scattering length $a_{12}$
 as well as for a fixed $a_{12}$ as  a function of  $N_1$ obtained by 
a variational, viz.  (\ref{gaussian_ansatz}), and a numerical solution of the
coupled NLS equations (\ref{GPEs1})-(\ref{GPEs2})
  using either an  LHY interaction or  a tunable  {\cite{tune}} three-body interaction $K_3$. 
The variation of the rms sizes of the binary balls, which is same for the two components in SMA,
as a function of $|a_{12}|$ and $N_1$ when only LHY interaction term is included in NLS equations (\ref{GPEs1})-(\ref{GPEs2})
is shown in figure \ref{fig20}(a) and figure \ref{fig20}(b), respectively. The same when only the  tunable
three-body interaction is considered is shown in figures \ref{fig20}(c) and (d). 
{
In figures \ref{fig20} the agreement between variational and numerical results is reasonable. }
 The tunable
$K_3$ is obtained using  (\ref{K3_estimate}) and gives $K_3\sim 10^{-39}-10^{-38}$m$^6$/s.
We find that either by using only LHY interaction or by using only three-body interaction,  binary 
quantum balls of similar sizes can be obtained as is illustrated in figures \ref{fig20}(a)-(d).
Hence for a proper description of the binary ball both the LHY and three-body interactions should be included in the Lagrangian. 
{
The values of $K_3$ used in figures \ref{fig5} and \ref{fig20} are a bit large. Nevertheless, there are suggestions for experimentally managing the  value of $K_3$ by external electromagnetic interactions \cite{tune}, which might provide a way to achieve such large values of $K_3$.  However, if we take a smaller value of $K_3$, a self-bound state will also emerge,  which will have a much smaller size.  }

}

\subsection{Moving quantum balls}

\label{IIIB}

\begin{figure}[t]
\begin{center} 
\includegraphics[trim = 0.25cm 0cm .25cm 0cm, clip,width=0.49\linewidth,clip]{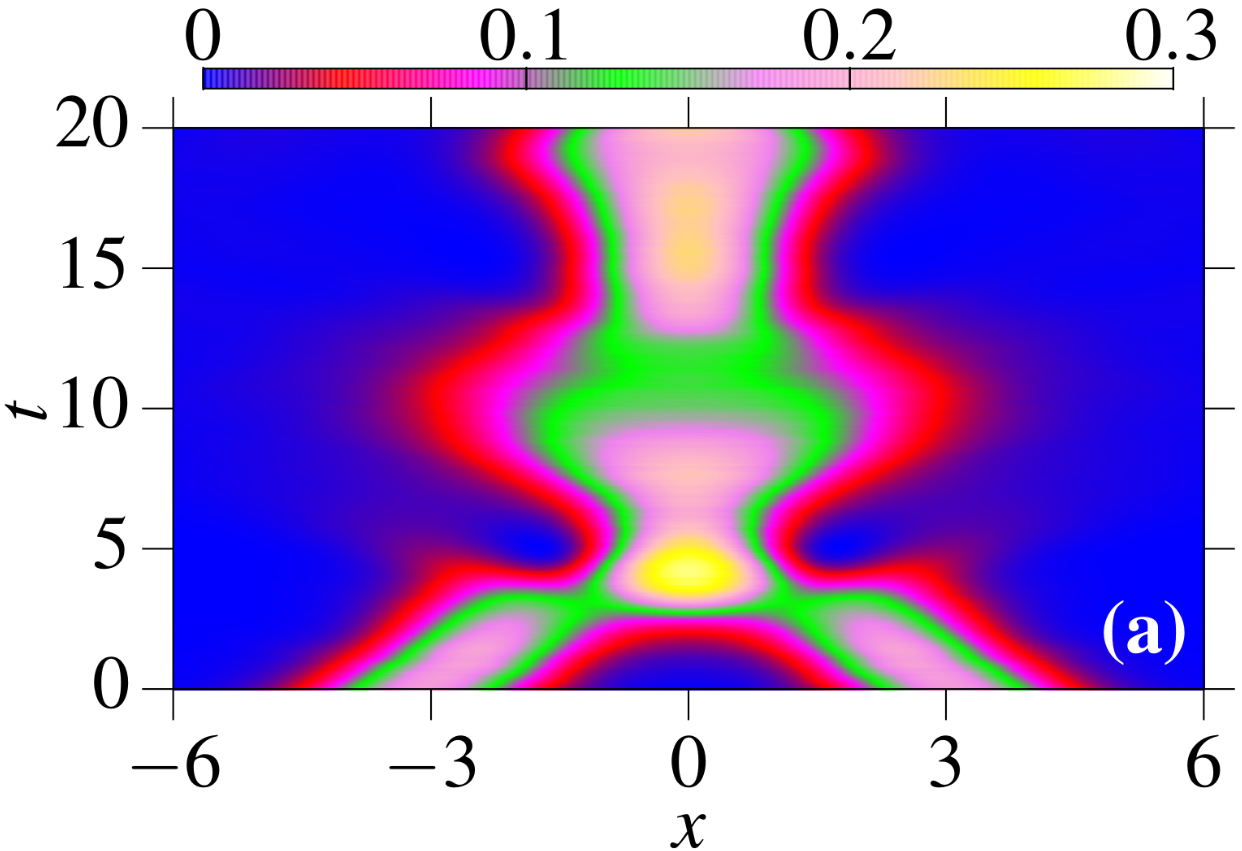}
\includegraphics[trim = 0.25cm 0cm .25cm 0cm, clip,width=0.49\linewidth,clip]{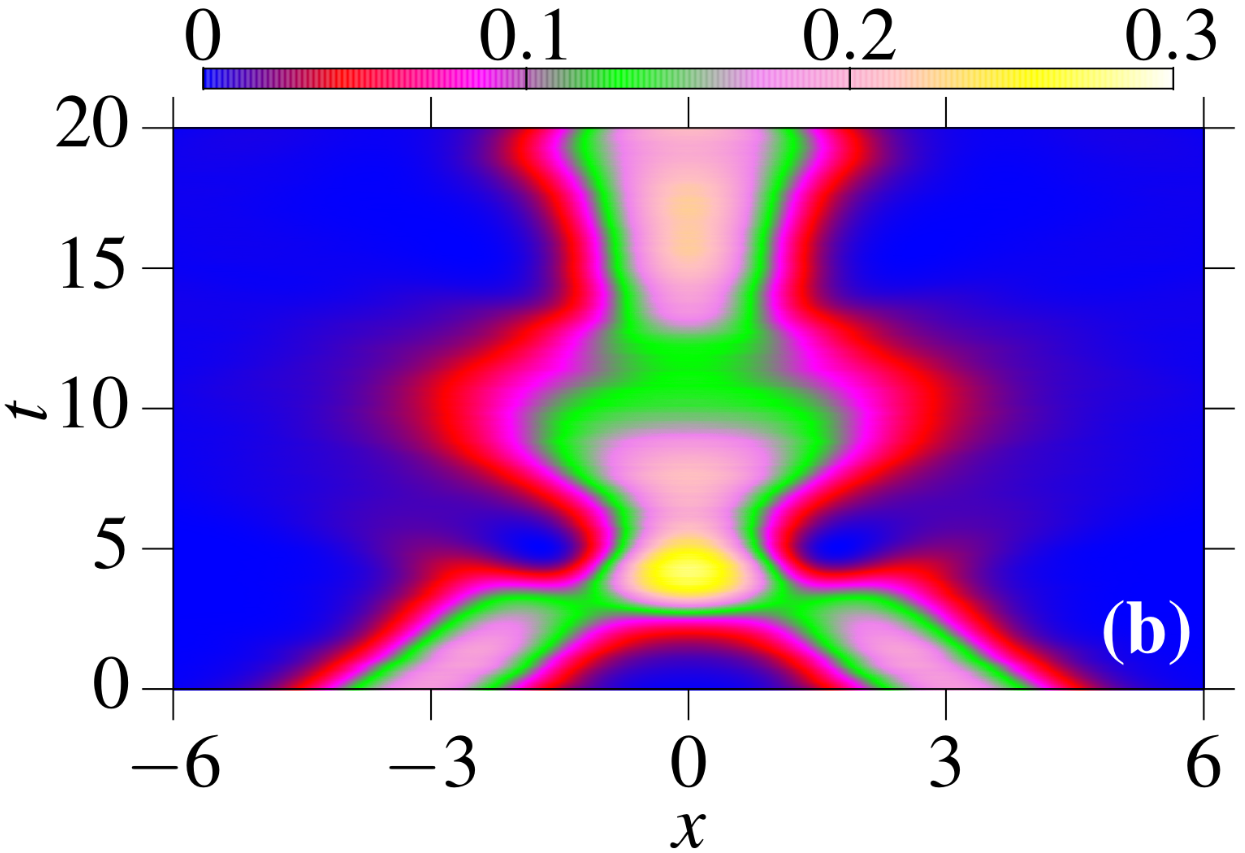} 
\caption{The frontal collision of two binary balls of figure \ref{fig2}(a) initially placed at 
$x_0=\pm 3.12$ moving in opposite directions  along $x$ axis with velocity $v_0=\pm 0.5$ through a two 
dimensional contour plot of density $\sum_{j=1,2}n_j(x,y=0,z=0,t)$
in the $x-t$ plane using (a) $K_3=1.8\times 10^{-41}$ m$^6$/s and (b) $K_3=1.8(1-{\mathrm i})\times 10^{-41}$ m$^6$/s, respectively.  }
\label{fig7}
\end{center}
\end{figure}

\begin{figure}[t]
\begin{center} 
\includegraphics[trim = 0.cm 0cm 0.cm 0cm, clip,width=0.49\linewidth,clip]{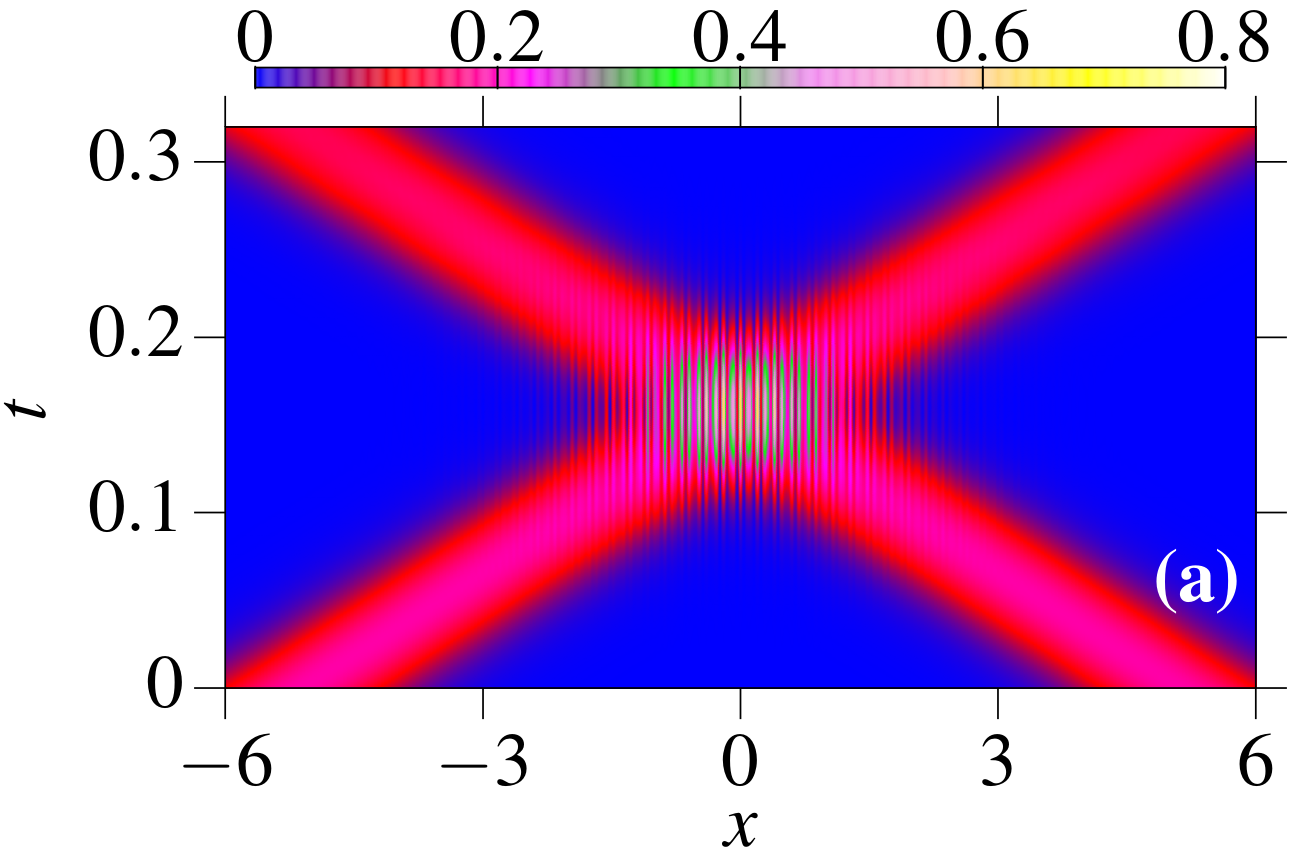}
\includegraphics[trim = 0.cm 0cm 0.cm 0cm, clip,width=0.49\linewidth,clip]{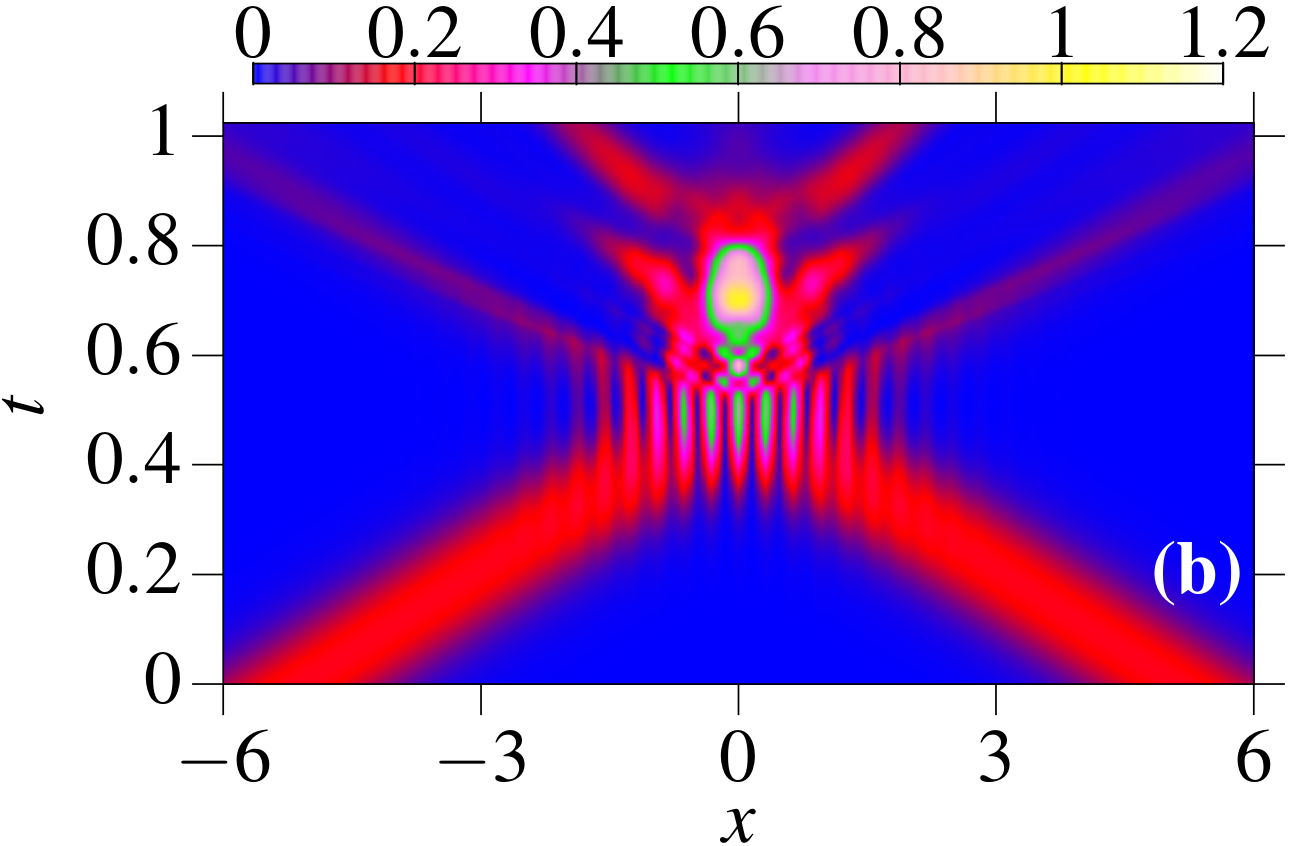} 
\caption{The frontal collision of two binary balls of figure \ref{fig2}(a) initially placed at 
$x_0=\pm 5.12$ moving in opposite directions  along $x$ axis with velocity $v_0=\pm 32$ and $\pm 10$ through a two 
dimensional contour plot of density $\sum_jn_j(x,y=0,z=0,t)$
in the $x-t$ plane. }
\label{fig8}
\end{center}
\end{figure}

To make the quantum balls move with a velocity, say $v_0$ along $x$ axis, we multiply
the stationary ground-state wave function obtained with imaginary time propagation  by $e^{i v_0x}$,
and evolve this solution in real time.  
We consider the binary  $^{87}$Rb  ball shown in {figure} \ref{fig2} (a) to study the collision
dynamics. To this end, we place two binary balls of figure \ref{fig2} (a) at 
$x=\pm x_0$ and attribute velocities $v=\pm v_0$ to them so that they collide frontally at 
$x=0$. We find that the head-on collision is essentially elastic at very large velocities 
with two balls emerging after collision with no visible deformation in shape.
    However, as
the speed of the colliding binary balls is decreased there is an increased deformation
in the shape  due to  collision.
 In the opposite extreme of very small velocities,  the collision is highly inelastic, and the identity of the two 
binary balls is lost through the formation of a larger binary ball $-$ a breather or a molecule $-$   in an excited state  
which stays at the origin ($x=0$) executing breathing oscillation.
The three-body interaction coefficient $K_3$ is actually complex with a negative imaginary part responsible for {the} loss of atoms due to 
three-body recombination. The imaginary part of $K_3$ in the  case of $^{87}$Rb atoms is quite small \cite{Tojo,tojo2}: 
$K_3=-1.8 {\mathrm i}\times 10^{-41}$ m$^6$/s. { In the real-time simulation of  collisions we consider an identical amount of the real part of $K_3$, so that  the resultant complex $K_3$ will be
 1.8 $({1-\mathrm i})\times 10^{-41}$ m$^6$/s.} However, such a small imaginary part of $K_3$  
at a low density of a quantum binary ball of few thousand atoms will lead to a small loss of atoms 
and would not destroy the  essentials of collision dynamics.       {
 However, the effect of the inclusion of the imaginary part of $K_3$ in the calculation of stationary quantum balls of Sec. \ref{IIIA} is insignificant. Hence we did not include an imaginary part of $K_3$ in that calculation.
 The effect of including an imaginary part of $K_3$ in collision dynamics} is shown in figures \ref{fig7}(a)-(b)
for the two binary  balls initially placed as $x_0=\pm 3.12$   moving with   speeds $v_0=\pm 0.5 $  in opposite directions employing 
$K_3=  1.8 \times 10^{-41}$ m$^6$/s and 
$K_3= 1.8 (1-i)\times 10^{-41}$ m$^6$/s, respectively. In both cases the two binary balls combine to form an excited  breather  
which oscillates at $x=0$. The difference in density in plots of figure \ref{fig7} (a) and (b) is not noticeable showing that the recombination loss is negligibly small.    
In figure \ref{fig8}(a)-(b), we illustrate similar collision of two binary balls at 
larger velocities $v_0=32$ and 10, respectively. For $v_0=\pm 32$, clean tracks of the binary balls are found to emerge after collision 
in the $x-t$ plane indicating the elastic nature of the collision. For $v_0=\pm 10$, the identity of the binary balls is lost after collision  as 
indicated by the diffused trails of the quantum balls after collision. 

\begin{figure}[t]
\begin{center}
\includegraphics[trim = 0cm 0cm 0cm -1cm, clip,width=0.3\linewidth,clip]{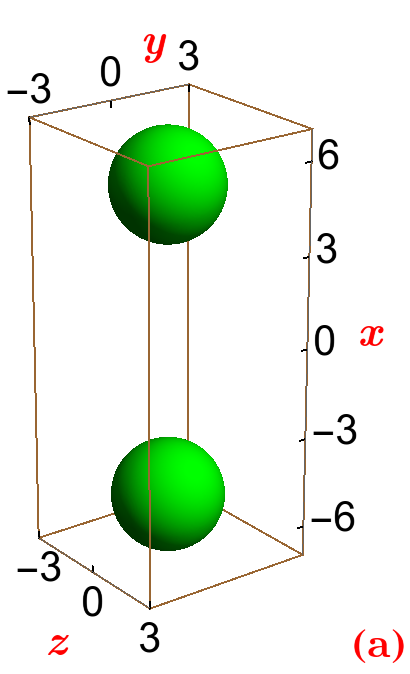}
\includegraphics[trim = 0cm 0cm 0cm -1cm, clip,width=0.3\linewidth,clip]{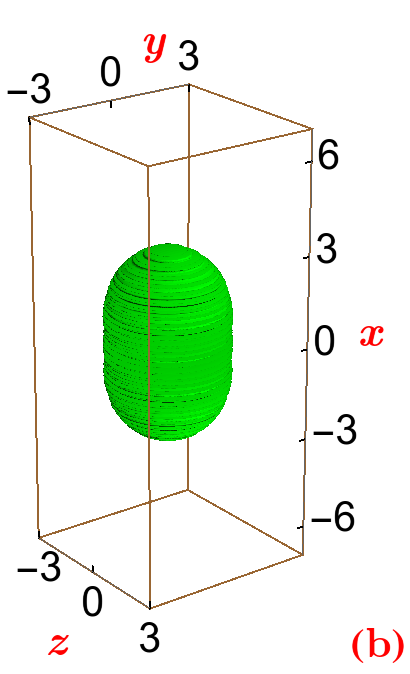}
\includegraphics[trim = 0cm 0cm 0cm -1cm, clip,width=0.3\linewidth,clip]{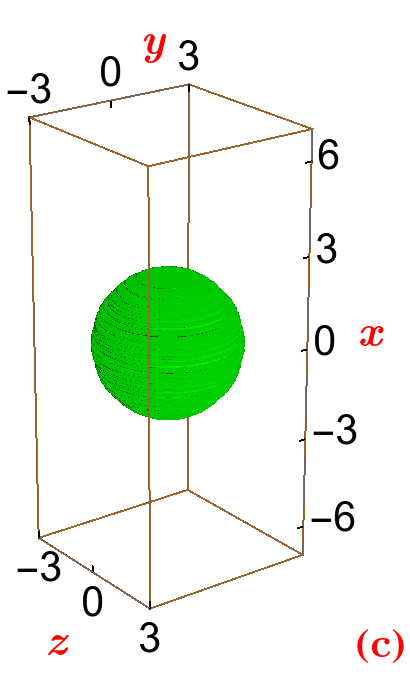}
\includegraphics[trim = 0cm 0cm 0cm -1cm, clip,width=0.3\linewidth,clip]{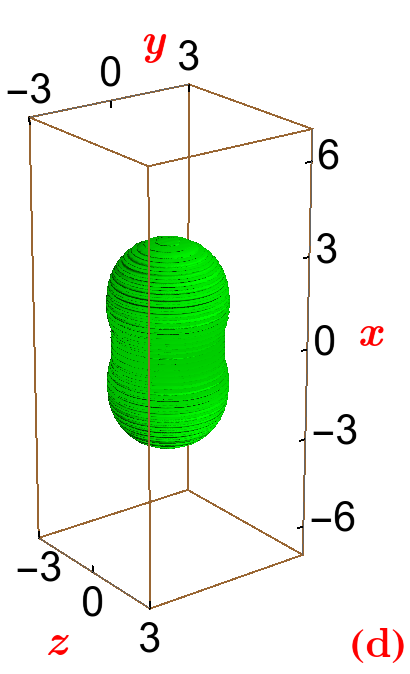}
\includegraphics[trim = 0cm 0cm 0cm -1cm, clip,width=0.3\linewidth,clip]{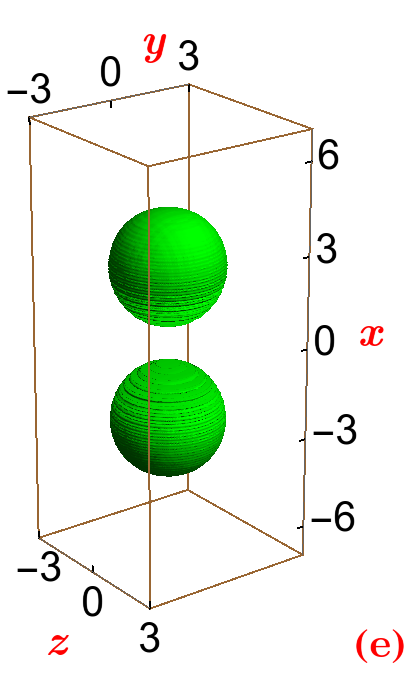}
\includegraphics[trim = 0cm 0cm 0cm -1cm, clip,width=0.3\linewidth,clip]{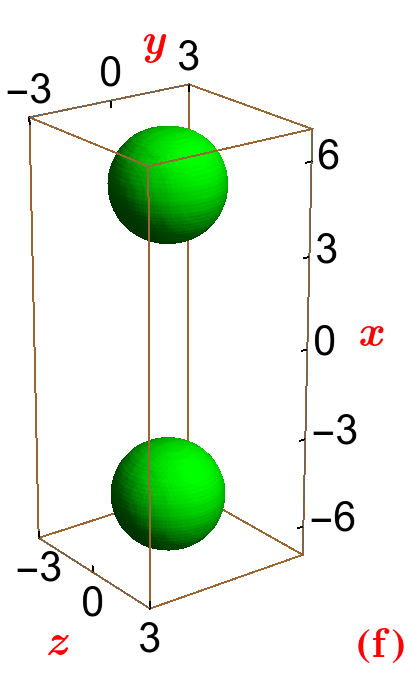}
\caption{Isodensity contours with  contour density  of $0.02$ of the two quantum balls 
corresponding to dynamics shown in {figure \ref{fig8}(a)} but calculated   
 with $K_3=1.8\times 10^{-41}(1-100\mathrm i)$ 
by 
real-time simulation 
at (a) $t =0$, (b) $t = 0.12$,
(c) $t= 0.16$ (d) $t= 0.21$, (e) $t=0.24$, and (f) $t=0.32$. }
\label{fig9}
\end{center}
\end{figure}

The quasi-elastic nature of collision of figure \ref{fig8}(a)  for $v=\pm 32$ is illustrated further by snapshots of 3D 
 {isodensity} contours      at different times, $t=0,0.12, 0.16, 0.21, 0.24$ and 0.32,
during the collision in figure \ref{fig9}. In case of many atoms the imaginary part of the three-body interaction coefficient $K_3$ 
has a larger value corresponding to a larger recombination loss { \cite{rate}.} To see that a larger value of the recombination loss does not destroy the elastic nature of the collision, 
in this simulation, we consider a $K_3$ with a large imaginary part: $K_3=  1.8\times 10^{-41}(1-100{\mathrm i}) $ m$^6$/s. 
{
The contour plot of the total density $\sum_j n_j(x,y=0,z=0,t)$ in $x$-$t$ plane in this
case is indistinguishable from that in  Fig. \ref{fig8}(a) due to a still smaller recombination loss over a period of
$t = 0.32$}. 
In (c) the two balls have fully overlapped and 
formed a unstable larger ball. In (b) and (d) the overlap is partial.  The final (f) and initial (a) snapshots of the colliding  
binary balls look quite similar  indicating the elastic nature of the collision.

\section{Summary and discussion}
\label{IV}

We  studied the formation of a binary BEC quantum ball of two hyperfine states 
of $^{87}$Rb atoms  for intra-species repulsion and inter-species attraction in 
the presence of beyond-mean-field LHY and complex three-body interactions and 
demonstrate that the LHY and three-body  interactions play similar role in the formation of a binary quantum ball. {
 Nevertheless, the strength of the three-body interaction $K_3$ has to be increased, if the size of the quantum ball stabilized only by the LHY interaction  is required to be equal to the same of a quantum ball stabilized only by the three-body interaction.  On the other hand, if a smaller value of $K_3$ is employed the size of the quantum ball will be much smaller.} Hence in a comprehensive treatment   of the binary BEC quantum ball both these interactions should be included.   A similar conclusion was reached on the treatment of a binary boson-fermion quantum ball \cite{arxiv},  

In Sec. \ref{II} we  derived a set of coupled NLS equations for the binary BEC with two- and three-body interactions 
and beyond-mean-field LHY interaction. For the formation of a stable binary quantum ball the inter-species interaction 
is always taken to  be attractive and intra-species interaction {
repulsive}. The three-body interaction is taken to be 
repulsive  with an imaginary part responsible for recombination loss.  To make the binary NLS equations analytically tractable, 
we consider a single-mode approximation (SMA) to it valid under some simplifying assumptions on the scattering lengths and 
number of atoms involved. An analytical Gaussian variational approximation was developed in the SMA. In Sec. \ref{IIIA}, {we} 
performed a numerical solution to the NLS equations  in the SMA by imaginary-time simulation and compared the results for stationary binary balls 
with the corresponding results of variational approximation. We also considered a numerical solution to the binary NLS {equations} beyond SMA, 
where there is no analytic variational approximation for a comparison. In Sec. \ref{IIIB}, the results for collision dynamics obtained by real-time simulation are presented.  
The elastic collision of two binary balls is found to be elastic at large velocities with practically no deformation of the emerging balls after collision. The collision is inelastic at smaller velocities and at very low velocities the two binary balls combine to form a larger binary ball in an excited state $-$ a breather or a binary-ball molecule $-$ which executes breathing oscillation.

After the completion of this study we came to know about a similar investigation \cite{gui}.

  \section*{Acknowledgments}

S.K.A acknowledges the support by the Funda\c c\~ao de Amparo \`a Pesquisa do Estado de
S\~ao Paulo (Brazil) under project 2016/01343-7 and also by
the Conselho Nacional de Desenvolvimento Cient\'ifico e Tecnol\'ogico (Brazil) under project 303280/2014-0.

\end{document}